\pdfoutput=1
\documentclass[11pt,twoside,a4paper,cmspaper,final,collab]{cms-tdr}
\def\svnVersion{5f359a3-D}\def\svnDate{2023/10/26}\def\cmsCernNoTag{CERN-EP-2022-245}\def\cmsCernDate{\today}\def\cmsMessage{Published in the European Physical Journal C as \href{http://dx.doi.org/10.1140/epjc/s10052-023-12050-4}{\doi{10.1140/epjc/s10052-023-12050-4}.}}
\begin{document}\cmsNoteHeader{TOP-20-008}

\newcommand{\mtop}{\ensuremath{m_{\PQt}}\xspace}
\newcommand{\mtfit}{\ensuremath{m_{\PQt}^\text{fit}}\xspace}
\newcommand{\mtgen}{\ensuremath{m_{\PQt}^\text{gen}}\xspace}
\newcommand{\mtreco}{\ensuremath{m_{\PQt}^\text{reco}}\xspace}
\newcommand{\mWreco}{\ensuremath{m_{\PW}^\text{reco}}\xspace}
\newcommand{\mlb}{\ensuremath{m_{\ell\PQb}^\text{reco}}\xspace}
\newcommand{\Rb}{\ensuremath{R_{\PQb\PQq}^\text{reco}}\xspace}
\newcommand{\Pgof}{\ensuremath{P_\text{gof}}\xspace}
\newcommand{\vectheta}{\ensuremath{\vec{\theta}}\xspace}
\newcommand{\vecalpha}{\ensuremath{\vec{\alpha}}\xspace}
\newcommand{\vecbeta}{\ensuremath{\vec{\beta}}\xspace}
\newcommand{\datadisttext}{The uncertainty bands contain statistical uncertainties in the simulation, normalization uncertainties due to luminosity and cross section, jet energy correction uncertainties, and all uncertainties that are evaluated from event-based weights. A large part of the depicted uncertainties on the expected event yields are correlated. The lower panels show the ratio of data to the prediction.  A value of  $\mtgen  = 172.5 \GeV$ is used in the simulation.}
\newcommand{\PYTHIAeight}{\textsc{pythia}8\xspace}
\newcommand{\protprot}{\ensuremath{\Pp\Pp}\xspace}
\newlength\cmsTabSkip\setlength{\cmsTabSkip}{1ex}
\newcolumntype{d}[1]{D{.}{.}{#1}}
\newcolumntype{L}{>{$}l<{$}}
\newcolumntype{R}{>{$}r<{$}}
\newcolumntype{C}{>{$}c<{$}}

\ifthenelse{\boolean{cms@external}}{\providecommand{\cmsLeft}{upper\xspace}}{\providecommand{\cmsLeft}{left\xspace}}
\ifthenelse{\boolean{cms@external}}{\providecommand{\cmsRight}{lower\xspace}}{\providecommand{\cmsRight}{right\xspace}} 
\ifthenelse{\boolean{cms@external}}{\providecommand{\CMSLeft}{Upper\xspace}}{\providecommand{\CMSLeft}{Left\xspace}}
\ifthenelse{\boolean{cms@external}}{\providecommand{\CMSRight}{Lower\xspace}}{\providecommand{\CMSRight}{Right\xspace}} 

\cmsNoteHeader{TOP-20-008}
\title{Measurement of the top quark mass using a profile likelihood approach with the lepton+jets final states in proton-proton collisions at \texorpdfstring{$\sqrt{s}=13\TeV$}{sqrt(s) = 13 TeV}}
\titlerunning{Measurement of the top quark mass using profile likelihood and lepton+jets at \texorpdfstring{$\sqrt{s}=13\TeV$}{sqrt(s) = 13 TeV}}
\abstract{
The mass of the top quark is measured in  36.3\fbinv of LHC proton-proton collision data collected with the CMS detector at $\sqrt{s}=13\TeV$. The measurement uses a sample of top quark pair candidate events containing one isolated electron or muon and at least four jets in the final state. 
For each event, the mass is reconstructed from a kinematic fit of the decay products to a top quark pair hypothesis. A  profile likelihood method is applied using up to four observables per event to extract the top quark mass. 
The top quark mass is measured to be $171.77\pm 0.37\GeV$. This approach significantly improves the precision over previous measurements.
\par
\vspace{2ex}
\textit{We dedicate this paper to the memory of our friend and colleague Thomas Ferbel, whose innovative work on precision measurements of the top quark mass laid the foundation for this publication.}}

\hypersetup{%
pdfauthor={CMS Collaboration},%
pdftitle={A profile likelihood approach to measure the top quark mass in the lepton+jets channel at  sqrt(s) = 13 TeV},%
pdfsubject={CMS},%
pdfkeywords={CMS, top, quark, mass, lepton, jets}} 

\date{\today}

\maketitle

\section{Introduction}
The top quark~\cite{Abe:1995hr,D0:1995jca} is the most massive fundamental particle and its mass, \mtop, is an important free parameter of the standard model (SM) of particle physics. 
 Because of its large Yukawa coupling, the top quark dominates the higher-order corrections to the
Higgs boson mass and  a precise determination  of \mtop sets strong constraints on the stability of the electroweak vacuum~\cite{HiggsStab,HiggsStab1}.
In addition, precise measurements of \mtop can be used to test the internal consistency of the SM~\cite{ALEPH:2013dgf,Baak:2012kk,Baak:2014ora}.

{\tolerance=800
At the CERN LHC, top quarks are produced predominantly in quark-antiquark pairs (\ttbar), which decay almost exclusively into a bottom (\PQb) quark and a \PW~boson.
Each \ttbar event can be classified by the subsequent decay of the \PW~bosons.
For this paper, the lepton+jets channel is analyzed, where  one \PW~boson decays hadronically, and the other leptonically.
Hence, the minimal final state consists of a muon or electron, at least four jets, and one undetected neutrino.
This includes events where a muon or electron from a tau lepton decay passes the selection criteria.
\par}

The mass of the top quark has been measured with increasing precision using the reconstructed invariant mass of different combinations of its decay products~\cite{PDG2020}.
The measurements by the Tevatron collaborations led to a combined value of $\mtop=174.30\pm0.65\GeV$~\cite{TevatronElectroweakWorkingGroup:2016lid}, while the ATLAS and CMS Collaborations measured $\mtop=172.69\pm0.48\GeV$~\cite{Aaboud:2018zbu} and $\mtop=172.44\pm0.48\GeV$~\cite{Khachatryan:2015hba}, respectively, from the combination of their most precise results at $\sqrt{s}=7$ and 8\TeV (Run~1).
The LHC measurements achieved a relative precision on \mtop of 0.28\%.
These analyses extract \mtop by comparing data directly to Monte Carlo simulations for different values of \mtop. 
An overview of the discussion of this mass definition and its relationship to a theoretically well-defined parameter is presented in Ref.~\cite{Hoang:2020iah}.

{\tolerance=800
In the lepton+jets channel, \mtop was measured by the CMS Collaboration with proton-proton (\protprot) collision data at $\sqrt{s} = 13\TeV$.
The result of $\mtop = 172.25\pm0.63\GeV$~\cite{Sirunyan:2018gqx} was extracted using the ideogram method~\cite{Abdallah:2008xh,Chatrchyan:2012cz}, which had previously been employed in Run~1~\cite{Khachatryan:2015hba}.
In contrast to the Run~1 analysis, in the analysis of $\sqrt{s} = 13\TeV$ data, the renormalization and factorization scales in the matrix-element (ME) calculation and the scales in the initial- and final-state parton showers (PS) were varied separately, in order to evaluate the corresponding systematic uncertainties.
In addition, the impacts of extended models of color reconnection (CR) were evaluated. 
These models were not available for the Run~1 measurements and their inclusion resulted in an increase in the systematic uncertainty~\cite{Sirunyan:2018gqx}. 
\par}

{\tolerance=800
In this paper, we use a new mass extraction method on the same data, corresponding to 36.3\fbinv,  that were used in Ref.~\cite{Sirunyan:2018gqx}.
In addition to developments in the mass extraction technique,  the reconstruction and calibration of the analyzed data have been improved, and updated simulations are used.
For example, the underlying event tune CP5~\cite{CMS:2019csb} and the jet flavor tagger \textsc{DeepJet}~\cite{Bols:2020bkb} were not available in the former analysis on the data.
The new analysis employs a kinematic fit of the decay products to a \ttbar  hypothesis.
For each event, the best matching assignment of the jets to the decay products is used.
A profile likelihood fit is performed using up to four different observables for events that are well reconstructed by the kinematic fit and one observable for the remaining events. 
The additional observables are used to constrain the main sources of systematic uncertainty.
The  model for the likelihood incorporates the effects of variations of these sources, represented by nuisance parameters based on simulation, as well as the finite size of the simulated samples.
This reduces the influence of systematic uncertainties in the measurement.
Tabulated results are provided in the HEPData record for this analysis~\cite{hepdata}.
\par}

\section{The CMS detector and event reconstruction}
The central feature of the CMS apparatus is a superconducting solenoid of 6\unit{m} internal diameter, which provides a magnetic field of 3.8\unit{T}. Within the solenoid volume are a silicon pixel and strip tracker, a lead tungstate crystal electromagnetic calorimeter (ECAL), and a brass and scintillator hadron calorimeter (HCAL), each composed of a barrel and two endcap sections. Forward calorimeters extend the pseudorapidity ($\eta$) coverage provided by the barrel and endcap detectors. Muons are measured in gas-ionization detectors embedded in the steel flux-return yoke outside the solenoid. A more detailed description of the CMS detector, together with a definition of the coordinate system used and the relevant kinematic variables, can be found in Ref.~\cite{CMS:2008xjf}. 
 
 {\tolerance=800
 The primary vertex is taken to be the vertex corresponding to the hardest scattering in the event, evaluated using tracking information alone, as described in Section 9.4.1 of Ref.~\cite{CMS-TDR-15-02}.
The particle-flow (PF) algorithm~\cite{CMS:2017yfk} aims to reconstruct and identify each individual particle in an event, with an optimized combination of information from the various elements of the CMS detector. 
The energy of photons is obtained from the ECAL measurement. 
The energy of electrons is determined from a combination of the electron momentum at the primary interaction vertex as determined by the tracker, the energy of the corresponding ECAL cluster, and the energy sum of all bremsstrahlung photons spatially compatible with originating from the electron track. 
The energy of muons is obtained from the curvature of the corresponding track. 
The energy of charged hadrons is determined from a combination of their momentum measured in the tracker and the matching ECAL and HCAL energy deposits, corrected for the response function of the calorimeters to hadronic showers. 
Finally, the energy of neutral hadrons is obtained from the corresponding corrected ECAL and HCAL energy deposits.
\par}

Jets are clustered from PF candidates using the anti-\kt algorithm with a distance parameter of 0.4~\cite{Cacciari:2008gp,Cacciari:2011ma}.
The jet momentum is determined as the vectorial sum of all particle momenta in the jet, and is found from simulation to be, on average, within 5 to 10\% of the true momentum over the whole transverse momentum (\pt) spectrum and detector acceptance. Additional \protprot interactions within the same or nearby bunch crossings  (pileup) can contribute additional tracks and calorimetric energy depositions, increasing the apparent jet momentum. To mitigate this effect, tracks identified as originating from pileup vertices are discarded and an offset correction is applied to correct for remaining contributions. Jet energy corrections are derived from simulation studies so that the average measured energy of jets becomes identical to that of particle level jets. In situ measurements of the momentum balance in dijet, photon+jet, $\PZ+$jet, and multijet events are used to determine any residual differences between the jet energy scale in data and in simulation, and appropriate corrections are made~\cite{Khachatryan:2016kdb}. Additional selection criteria are applied to each jet to remove jets potentially dominated by instrumental effects or reconstruction failures.
The jet energy resolution amounts typically to 15--20\% at 30\GeV, 10\% at 100\GeV, and 5\% at 1\TeV~\cite{Khachatryan:2016kdb}.
Jets originating from \PQb quarks are identified using the \textsc{DeepJet} algorithm~\cite{Sirunyan:2017ezt, Bols:2020bkb,CMS-DP-2018-058}.
This has an efficiency of approximately 78\%, at a misidentification probability of 1\% for light-quark and gluon jets and 12\% for charm-quark jets~\cite{Bols:2020bkb,CMS-DP-2018-058}.

The missing transverse momentum vector, \ptvecmiss, is computed as the negative vector sum of the transverse momenta of all the PF candidates in an event, and its magnitude is denoted as \ptmiss~\cite{Sirunyan:2019kia}. The \ptvecmiss is modified to account for corrections to the energy scale of the reconstructed jets in the event.

The momentum resolution for electrons with $\pt \approx 45\GeV$ from $\PZ \to \Pe \Pe$ decays ranges from 1.6 to 5.0\%. It is generally better in the barrel region than in the endcaps, and also depends on the bremsstrahlung energy emitted by the electron as it traverses the material in front of the ECAL~\cite{CMS:2020uim,CMS-DP-2020-021}.

Muons are measured in the pseudorapidity range $\abs{\eta} < 2.4$, with detection planes made using three technologies: drift tubes, cathode strip chambers, and resistive plate chambers. 
Matching muons to tracks measured in the silicon tracker results in a relative transverse momentum resolution, for muons with \pt up to 100\GeV, of 1\% in the barrel and 3\% in the endcaps. The \pt resolution in the barrel is better than 7\% for muons with \pt up to 1\TeV~\cite{Sirunyan:2018fpa}.

\section{Data samples and event selection}\label{sec:evtsel}
{\tolerance=800
The analyzed data sample has been collected with the CMS detector in 2016 at a center-of-mass energy $\sqrt{s} = 13\TeV$. 
It corresponds to an integrated luminosity of 36.3\fbinv~\cite{CMS:2021xjt}.
Events are required to pass a single-electron trigger with a \pt threshold for isolated electrons of 27\GeV or a single-muon trigger with a minimum threshold on the \pt of an isolated muon of 24\GeV~\cite{Khachatryan:2016bia}.
\par}

{\tolerance=1400
Simulated \ttbar  signal events are generated with the \POWHEG~v2 ME generator~\cite{Nason:2004rx,Frixione:2007vw,Alioli:2010xd}, \textsc{pythia}8.219 PS~\cite{Sjostrand:2014zea}, and use the CP5 underlying event tune~\cite{CMS:2019csb} with  top quark mass values, \mtgen, of 169.5, 172.5, and 175.5\GeV. 
To model parton distribution functions (PDFs), the  NNPDF3.1 next-to-next-to-leading order (NNLO) set~\cite{Butterworth:2015oua,Ball:2017nwa}  is used with the strong coupling constant set to $\alpS = 0.118$.
The various background samples are simulated with the same ME generators and matching techniques~\cite{Alioli:2009je, Re:2010bp, Alwall:2014hca,Alwall:2007fs,Frederix:2012ps} as in Ref.~\cite{Sirunyan:2018gqx}.
The background processes are $\PW/\PZ$+jets, single-top, diboson, and events composed uniquely of jets produced through the strong interaction, referred to as quantum chromodynamics (QCD) multijet events.
The PS simulation and hadronization is performed with \PYTHIAeight, using the CUETP8M1 tune~\cite{Skands:2014pea}.
\par}

{\tolerance=2000
All of the simulated samples are processed through a full simulation of the CMS detector based on \GEANTfour~\cite{Agostinelli:2002hh} and are normalized to their predicted cross section described in Refs.~\cite{Czakon:2011xx,Li:2012wna,Aliev:2010zk,Kant:2014oha}.
The effects of pileup are included in the simulation and the pileup distribution in simulation is weighted to match the pileup in the data. 
The jet energy response and resolution in simulated events are corrected to match the data~\cite{Khachatryan:2016kdb}. In addition, the \PQb-jet identification (\PQb~tagging) efficiency and misidentification rate~\cite{Sirunyan:2017ezt}, and the lepton trigger and reconstruction efficiencies are corrected in simulation~\cite{CMS:2020uim, Sirunyan:2018fpa}.
\par}

Events are selected with exactly one isolated electron (muon) with $\pt> 29$ (26)\GeV and $\abs{\eta}< 2.4$ that is separated from PF jet candidates with $\Delta R=\sqrt{\smash[b]{{(\Delta \eta)}^2+{(\Delta \phi)}^2}} >0.3$ (0.4), where $\Delta \eta$ and $\Delta \phi$ are the differences in pseudorapidity and azimuth (in radians) between the jet and lepton candidate.
The four leading jet candidates in each event are required to have $\pt>30\GeV$ and $\abs{\eta}<2.4$.
Only these four jets are used in further reconstruction.
Exactly two \PQb-tagged jets are required among the four selected jets, yielding 287\,842 (451\,618) candidate events in the electron+jets (muon+jets) decay channel.

To check the compatibility of an event with the \ttbar hypothesis, and to improve the resolution of the reconstructed quantities, a kinematic fit~\cite{Abbott:1998dc} is performed.
For each event, the inputs to the algorithm are the momenta of the lepton and of the four leading jets, \ptvecmiss, and the resolutions of these variables.
The fit constrains these quantities to the hypothesis that two heavy particles of equal mass are produced, each one decaying to a \PQb quark and a \PW boson, with the invariant mass of the latter constrained to 80.4\GeV.
The kinematic fit then minimizes $\chi^{2} \equiv (\mathbf{x}-\mathbf{x}^{m})^\mathrm{T}G(\mathbf{x}-\mathbf{x}^{m})$
where $\mathbf{x}^{m}$ and  $\mathbf{x}$ are the vectors of the measured and fitted momenta, respectively, and $G$ is the inverse covariance matrix, which is constructed from the uncertainties in the measured momenta. The masses are fixed to 5\GeV for the  \PQb quarks and to zero for the light quarks and leptons.
The two \PQb-tagged jets are candidates for the \PQb quarks in the \ttbar hypothesis, while the two jets that are not \PQb~tagged serve as candidates for the light quarks from the hadronically decaying \PW boson.
This leads to two possible parton-jet assignments, each with two solutions for the longitudinal component of the neutrino momentum, and four different permutations per event.
For simulated \ttbar events, the parton-jet assignments can be classified as correct permutations, wrong permutations, and unmatched permutations, where, in the latter case, at least one quark from the \ttbar decay is not unambiguously matched within a distance of $\Delta R <0.4$ to any of the four selected jets. 

\begin{figure}[ht]
\noindent \centering
\includegraphics[width=0.45\textwidth]{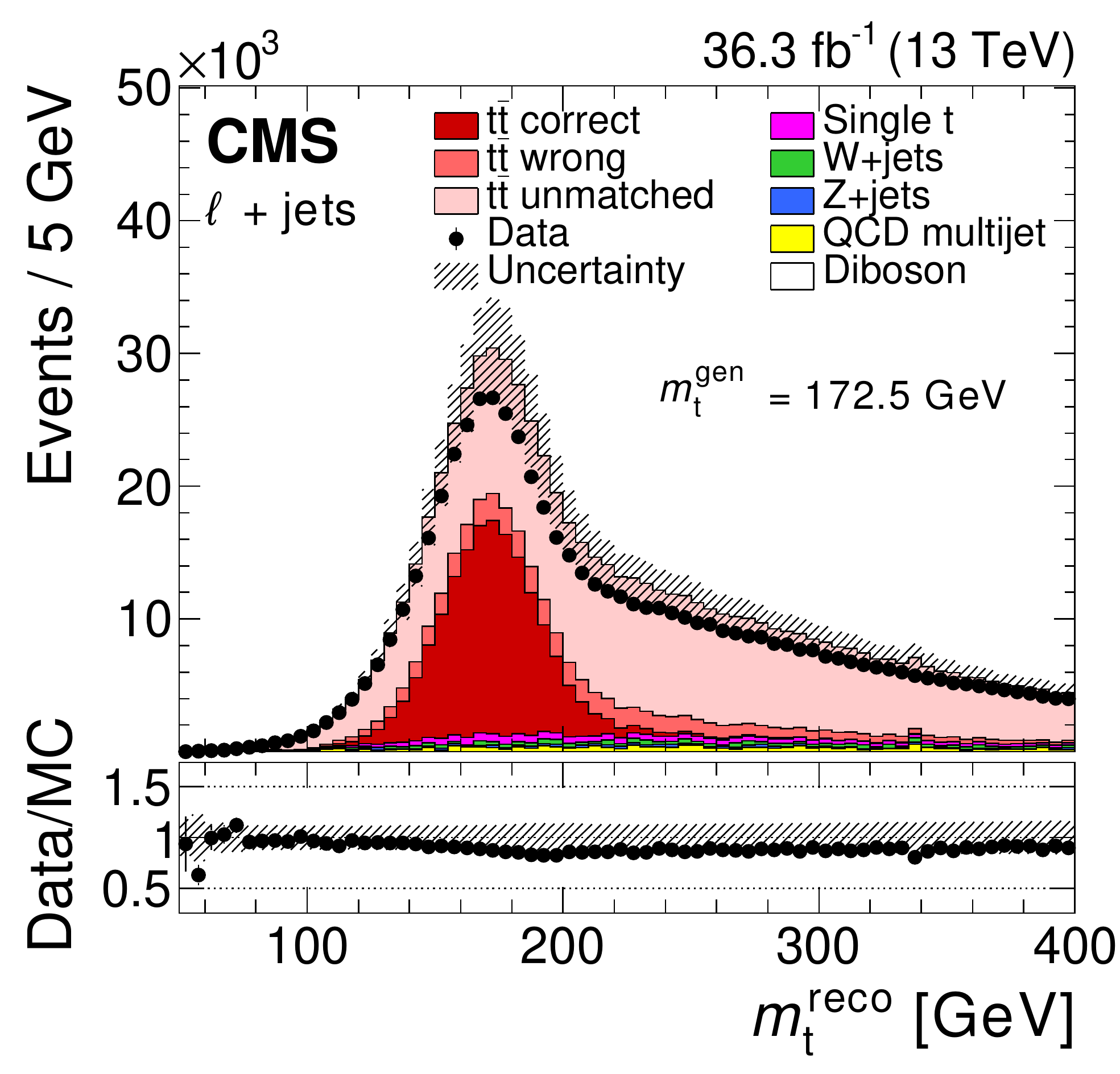}
\includegraphics[width=0.45\textwidth]{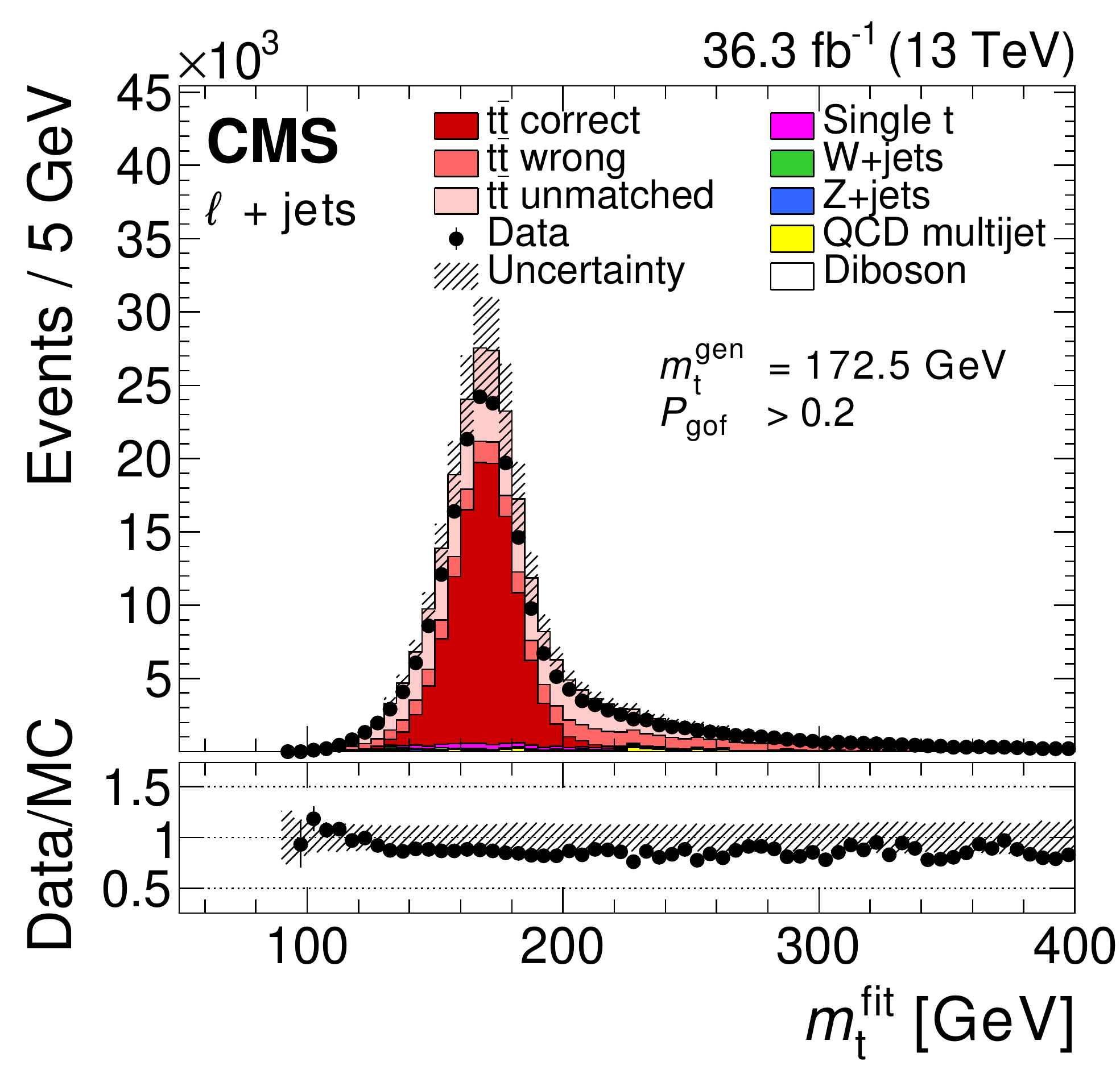}
\caption{\label{fig1}
The top quark mass distribution before (\cmsLeft) and after (\cmsRight) the $\Pgof > 0.2$ selection and the kinematic fit.  
For the simulated \ttbar events, the jet-parton assignments are classified as correct, wrong, and unmatched permutations, as described in the text.
\datadisttext}
\end{figure}

{\tolerance=1000
The goodness-of-fit probability, $\Pgof = \exp(-\chi^{2}/2)$, is used to determine the most likely parton-jet assignment.
For each event, the observables from the permutation with the highest \Pgof value are the input to the \mtop measurement.
In addition, the events are categorized as either $\Pgof < 0.2$ or $\Pgof > 0.2$, matching the value chosen in Ref.~\cite{Sirunyan:2018gqx}.
Requiring  $\Pgof > 0.2$ yields 87\,265 (140\,362) \ttbar candidate events in the electron+jets (muon+jets) decay channel and has  a predicted signal fraction of 95\%.
This selection improves the fraction of correctly reconstructed events from 20  to 47\%.
Figure~\ref{fig1} shows the distribution of the invariant mass of the hadronically decaying top quark candidate before (\mtreco) and after (\mtfit) the $\Pgof > 0.2$ selection and the kinematic fit.
A large part of the depicted uncertainties on the expected event yields are correlated.
Hence, the overall normalization of the simulation agrees within the uncertainties, although the simulation predicts 10\% more events in all distributions. 
For the final measurement, the simulation is normalized to the number of events observed in data.
\par}

\section{Observables and systematic uncertainties} \label{sec:obs}

Different observables are used per event based on its $\Pgof$ value. The cases and observables are listed in Table~\ref{tab:obs}.
\begin{table}[!htb]
  \topcaption{The overall list of different input histograms and their inclusion in a certain histogram set. A histogram marked with "$\times$" is included in a set (measurement). 
    \label{tab:obs}}
  \centering
  \begin{tabular}{llccccc}
     \multicolumn{2}{c}{Histogram} & \multicolumn{5}{c}{Set label}\\
    Observable & Category &  \textit{1D} &  \textit{2D} &  \textit{3D} &  \textit{4D} &  \textit{5D}   \\ \hline
    \mtfit  & $\Pgof > 0.2$ & $\times$ &$\times$ &$\times$ & $\times$ & $\times$  \\
    \mWreco &  $\Pgof > 0.2$ & & $\times$ & $\times$ & $\times$ & $\times$  \\ 
     \mlb &  $\Pgof < 0.2$ &  &   &  $\times$ &  $\times$  &  $\times$  \\
    $\mlb/\mtfit$  & $\Pgof > 0.2$ & & & & $\times$ & $\times$ \\ 
    \Rb &  $\Pgof > 0.2$ & &  &  &   & $\times$ \\ 
    \end{tabular}
\end{table}

\begin{figure}[ht]
\noindent \centering
\includegraphics[width=0.45\textwidth]{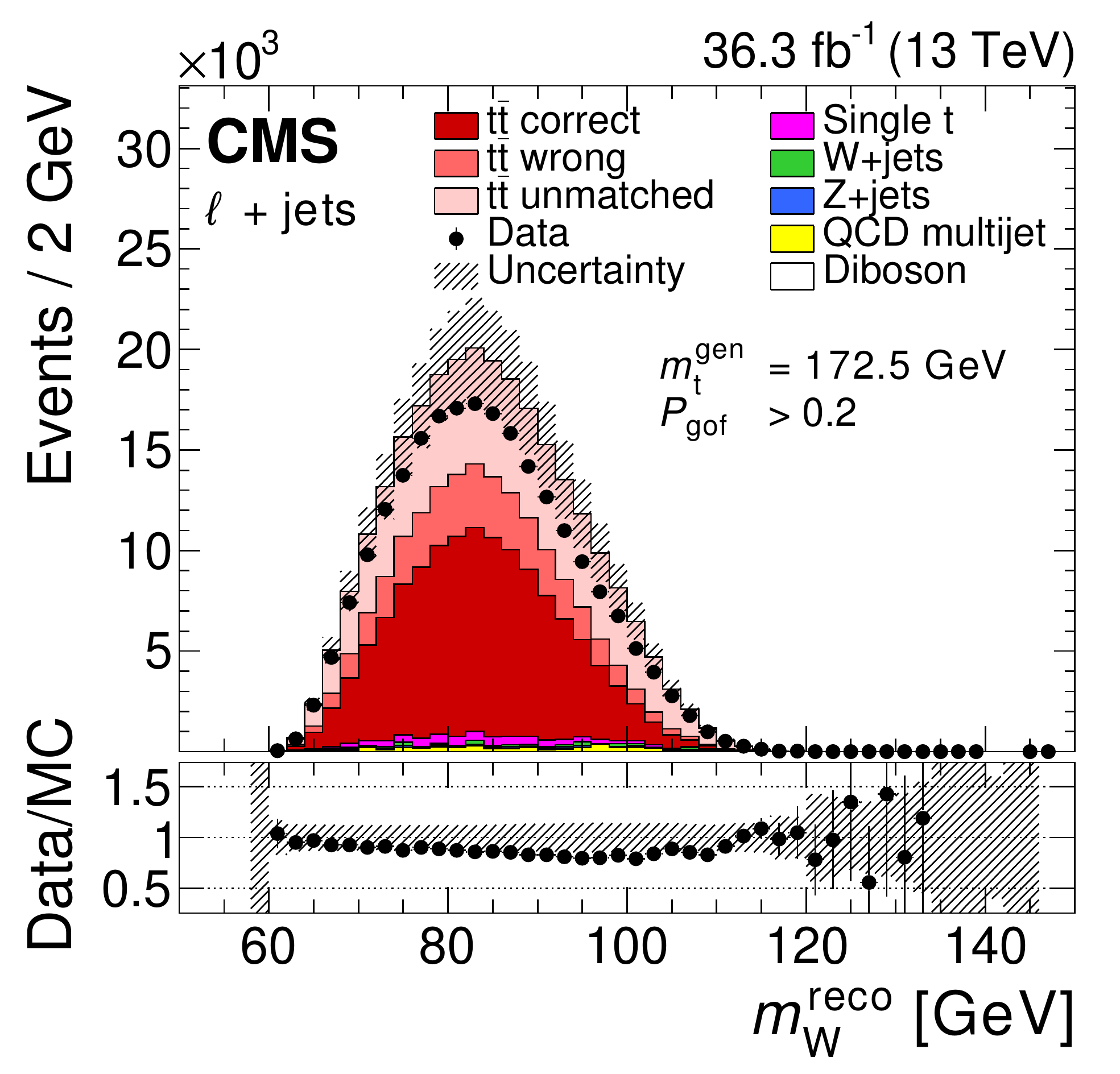}
\includegraphics[width=0.45\textwidth]{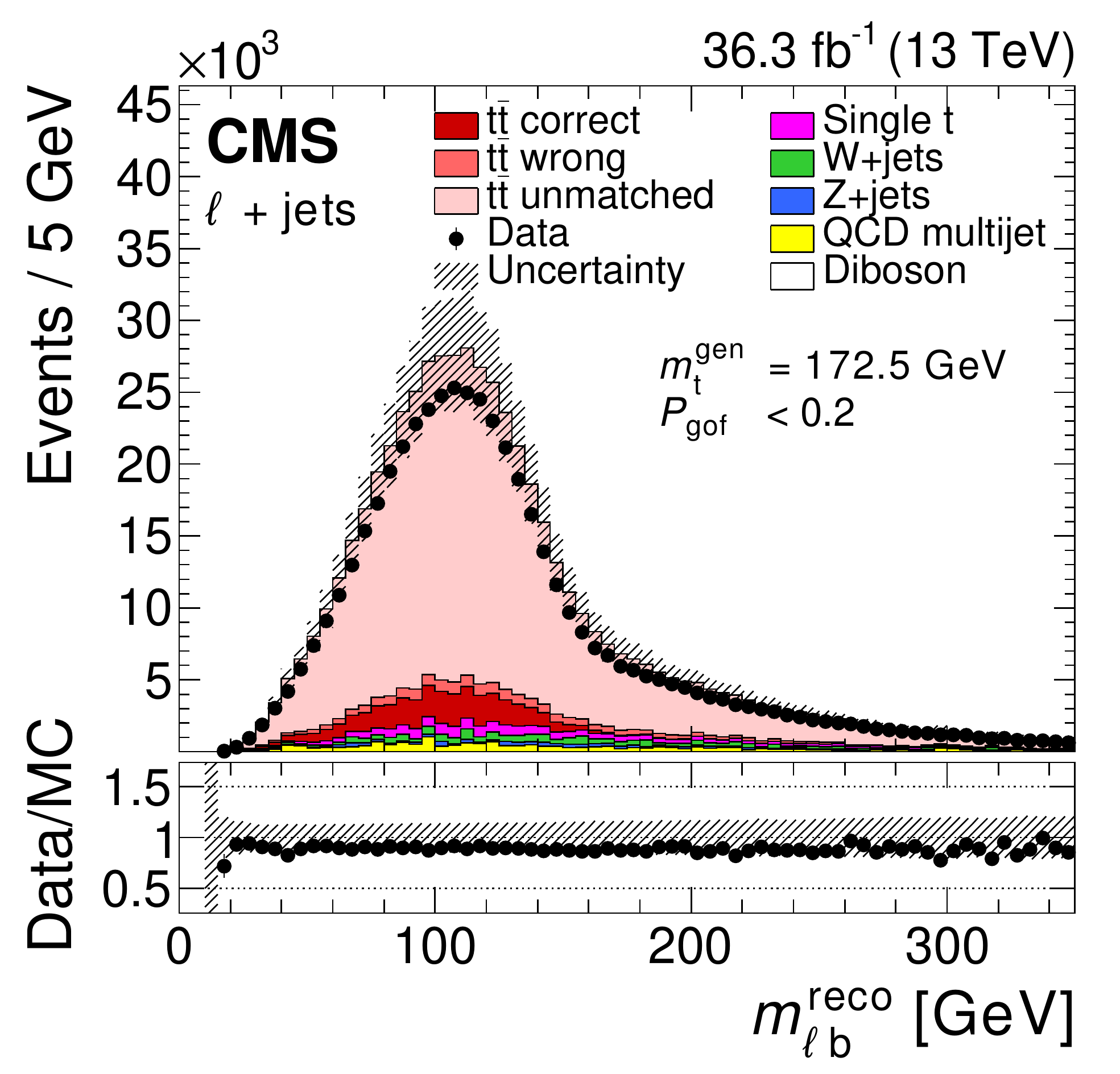}
\caption{\label{fig3}
The distributions of the reconstructed \PW~boson mass for the $\Pgof > 0.2$ category (\cmsLeft) and of the invariant mass of the lepton and the jet assigned to the semileptonic decaying top quark  for the  $\Pgof < 0.2$ category (\cmsRight). \datadisttext}
\end{figure}

For events with $\Pgof > 0.2$, the mass of the top quark candidates from the kinematic fit, \mtfit, shows a very strong dependence on \mtop and is the main observable in this analysis.
Only events with \mtfit values between 130 and 350 \GeV are used in the measurement. The high mass region is included to constrain the contribution of unmatched events to the peak.
For events with  $\Pgof < 0.2$, the invariant mass of the lepton and the \PQb-tagged jet assigned to the semileptonically decaying top quark, \mlb, is shown in Fig.~\ref{fig3} (\cmsRight).
For most \ttbar events, a low \Pgof  value is caused by assigning a wrong jet to the \PW boson candidate, while the correct \PQb-tagged jets are the candidates for the \PQb quarks in 60\% of these events.
Hence, \mlb preserves a good \mtop dependence and adds additional sensitivity to the measurement. 
The measurement is limited to \mlb values between 0 and 300\GeV.
While a similar observable has routinely been used in \mtop measurements in the dilepton channel~\cite{Aaboud:2016igd,Sirunyan:2018goh}, this is the first application of this observable in the lepton+jets channel.

\begin{figure}[ht]
\noindent \centering
\includegraphics[width=0.45\textwidth]{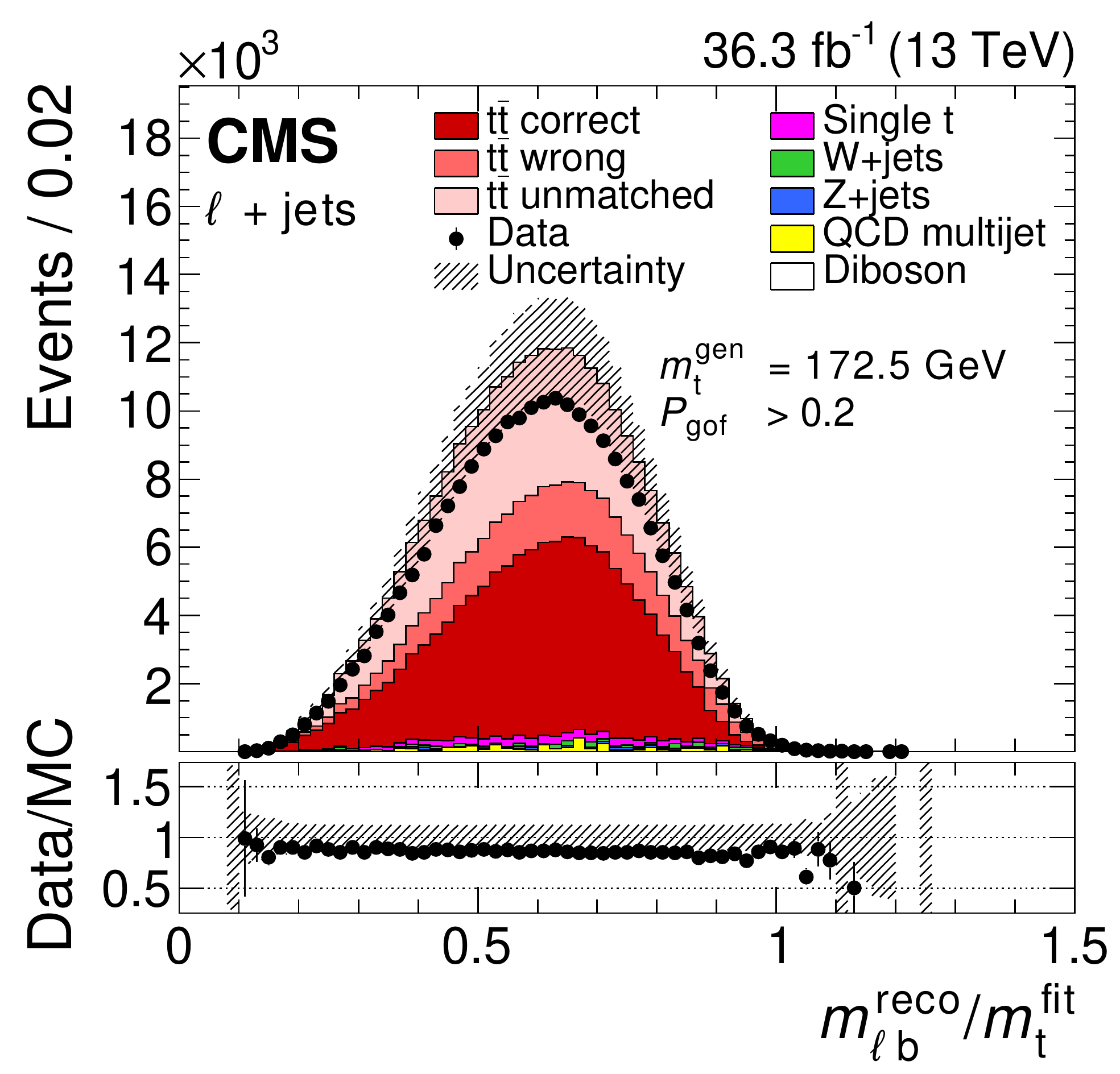}
\includegraphics[width=0.45\textwidth]{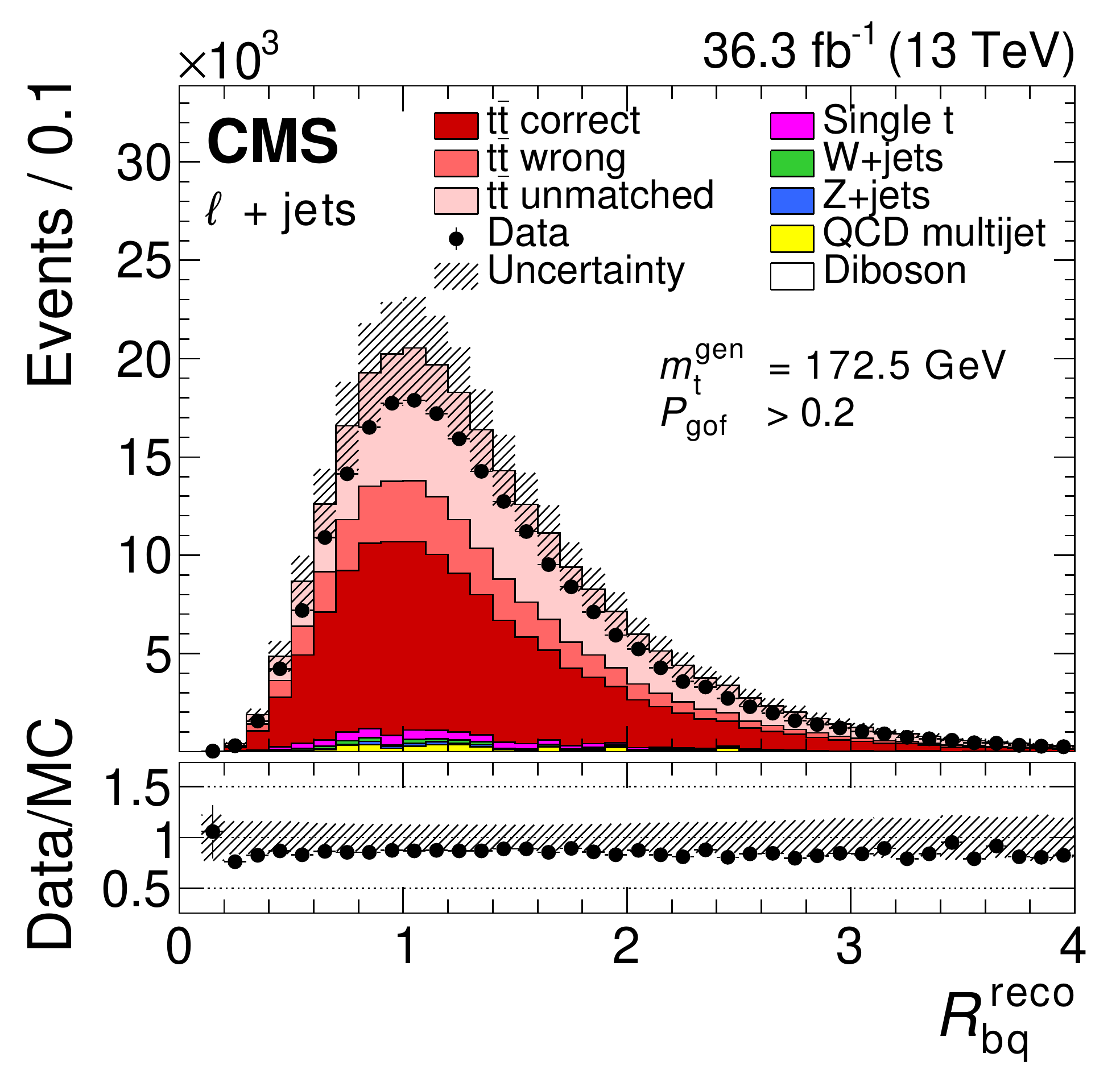}
\caption{\label{fig4}
The distributions of $\mlb/\mtfit$ (\cmsLeft) and of \Rb (\cmsRight), both for the $\Pgof > 0.2$ category. \datadisttext}
\end{figure}

Additional observables are used in parallel for the mass extraction to constrain systematic uncertainties. 
In previous analyses by the CMS Collaboration in the lepton+jets channel~\cite{Sirunyan:2018gqx,Khachatryan:2015hba}, the invariant mass of the two untagged jets before the kinematic fit, \mWreco, has been used together with \mtfit, mainly to reduce the uncertainty in the jet energy scale and the jet modeling. 
Its distribution is shown in  Fig.~\ref{fig3} (\cmsLeft) and the region between 63 and 110\GeV is used in the measurement.
As \mWreco is only sensitive to the energy scale and modeling of light flavor jets, two additional observables are employed to improve sensitivity to the scale and modeling of jets originating from \PQb  quarks. 
These are the ratio  $\mlb/\mtfit$, and the ratio of the scalar sum of the transverse momenta of the two \PQb-tagged jets ($\PQb 1$, $\PQb 2$),  and the two non-\PQb-tagged jets ($\PQq1$, $\PQq2$), $\Rb = (\pt^{\PQb1} + \pt^{\PQb2})/(\pt^{\PQq1}+\pt^{\PQq2})$. 
Their distributions are shown in Fig.~\ref{fig4}.
While \mtfit and \mWreco have been used  by the CMS Collaboration in previous analyses in the lepton+jets channel, \mlb,  $\mlb/\mtfit$, and \Rb are new additions.  
However,  \Rb has been used in the lepton+jets channel by the ATLAS Collaboration~\cite{Aad:2015nba,Aaboud:2018zbu}.

{\tolerance=5000
The distributions of the five observables are affected by uncertainties in the modeling and the reconstruction of the simulated events.
These sources of systematic uncertainties are nearly identical to those in the previous measurements~\cite{Sirunyan:2018gqx,Sirunyan:2018mlv}. 
The only difference is that we no longer include a systematic uncertainty related to the choice of ME generator as this uncertainty would overlap with the uncertainty we find from varying the default ME generator parameters.
The considered sources are summarized in the categories listed below.
\begin{itemize}
\item \textit{Method calibration:} In the previous measurements~\cite{Sirunyan:2018gqx,Sirunyan:2018mlv}, the limited size of the simulated samples for different values of \mtgen lead to an uncertainty in the calibration of the mass extraction method.  
In the new profile likelihood approach, the statistical uncertainty in the top quark mass dependence due to the limited sample size is included via nuisance parameters. 
\item \textit{Jet energy correction \textnormal{(JEC)}:} Jet energies are scaled up and down according to the \pt- and $\eta$-dependent data/simulation uncertainties~\cite{Khachatryan:2016kdb, CMS-DP-2020-019}.
Each of the 25 individual uncertainties in the jet energy corrections is represented by its own nuisance parameter.
\item \textit{Jet energy resolution \textnormal{(JER)}:} Since the JER measured in data is worse than in simulation,
the simulation is modified to correct for the difference~\cite{Khachatryan:2016kdb, CMS-DP-2020-019}.
The jet energy resolution in the simulation is varied up and down within the uncertainty.
The variation is evaluated independently for two $\abs{\eta_{\text{jet}}}$ regions, split at $\abs{\eta_{\text{jet}}}=1.93$.
\item \textit{\PQb~tagging:} The \pt-dependent uncertainty of the \PQb-tagging efficiencies and misidentification rates of the \textsc{DeepJet} tagger~\cite{Bols:2020bkb,CMS-DP-2018-058} are taken into account by reweighting the simulated events accordingly.
\item \textit{Pileup:} To estimate the uncertainty from the determination of the number of pileup events and the reweighting procedure, the inelastic \protprot cross section~\cite{Sirunyan:2018nqx} used in the determination is varied by $\pm$4.6\%.
\item \textit{Background \textnormal{(BG)}:} The main uncertainty in the background stems from the uncertainty in the measurements of the cross sections used in the normalization.
The normalization of the background samples is varied by $\pm$10\% for the single top quark samples~\cite{Khachatryan:2014iya,Sirunyan:2016cdg}, $\pm$30\% for the $\PW+$jets samples~\cite{Khachatryan:2016ipq}, $\pm$10\% for the $\PZ+$jets~\cite{Khachatryan:2016iob} and for the diboson samples~\cite{Khachatryan:2016tgp,Sirunyan:2017zjc}, and $\pm$100\% for the QCD multijet samples. The size of the variations is the same as in the previous measurement~\cite{Sirunyan:2018gqx} in this channel.
The uncertainty in the luminosity of 1.2\%~\cite{CMS:2021xjt} is negligible compared to these variations. 
\item \textit{Lepton scale factors \textnormal{(SFs)} and momentum scale:} The simulation-to-data scale factors  for the trigger, reconstruction, and selection efficiencies for electrons and muons are varied within their uncertainties. In addition,  the lepton energy in simulation is varied up and down within its uncertainty.
\item \textit{JEC flavor:} The difference between Lund string fragmentation and cluster fragmentation is evaluated by comparing \PYTHIA 6.422~\cite{Sjostrand:2006za} and \HERWIGpp~2.4~\cite{Bahr:2008pv}. 
The jet energy response is compared separately for each jet flavor~\cite{Khachatryan:2016kdb}.
\item \textit{\PQb-jet modeling  \textnormal{(bJES)}:} The uncertainty associated with the fragmentation of \PQb~quarks is split into four components.
  The Bowler--Lund fragmentation function is varied symmetrically within its uncertainties, as determined by the ALEPH and DELPHI
  Collaborations~\cite{DELPHI:2011aa,Heister:2001jg}.
  The difference between the default \PYTHIA setting and the center of the variations is included as an additional uncertainty.
  As an alternative model of the fragmentation into \PQb hadrons, the Peterson fragmentation function is used and the difference obtained relative to the Bowler--Lund fragmentation function is assigned as an uncertainty.
  The third uncertainty source taken into account is the semileptonic \PQb-hadron branching fraction, which is varied by
  ${-}0.45$ and ${+}0.77\%$, motivated by measurements of \PBz/\PBp decays and their corresponding uncertainties~\cite{PDG2020} and a comparison of the measured branching ratios to  \PYTHIA. 
\item \textit{PDF:}
        The default PDF set in the simulation, NNPDF3.1 NNLO~\cite{Butterworth:2015oua,Ball:2017nwa}, is replaced with the CT14 NNLO~\cite{Dulat:2015mca} and MMHT 2014 NNLO~\cite{Harland-Lang:2014zoa} PDFs via event weights.
        In addition, the default set  is varied with 100 Hessian eigenvectors~\cite{Ball:2017nwa} and the $\alpS$ value is changed to  0.117 and 0.119. 
        All described variations are evaluated for their impact on the measurement and the negligible variations are later omitted to reduce the number of nuisance parameters.
\item \textit{Renormalization and factorization scales:}
  The renormalization and factorization scales for the ME calculation are varied independently and simultaneously by factors of 2 and $1/2$.
  This is achieved by reweighting the simulated events. 
  The independent variations were checked and it was found to be sufficient to include only the simultaneous variations as a nuisance parameter.
\item \textit{ME to PS matching:} The matching of the \POWHEG ME calculations to the \PYTHIA PS is varied by shifting
  the parameter $h_{\text{damp}}=1.58^{+0.66}_{-0.59}$~\cite{CMS-PAS-TOP-16-021} within its uncertainty.
\item \textit{ISR and FSR:} For initial-state radiation (ISR) and final-state radiation (FSR), 32 decorrelated variations of the renormalization scale and nonsingular terms for each branching type ($\Pg\to \Pg\Pg$, $\Pg\to \PQq\PAQq$, $\PQq\to \PQq\Pg$, and  $\PX \to \PX\Pg$ with $\PX = \PQt\text{ or }\PQb$) are applied using event weights~\cite{Mrenna:2016sih}. 
  The scale variations correspond to  a change of the respective PS scale in \PYTHIA  by factors of 2 and $1/2$.
This approach is new compared to the previous analysis~\cite{Sirunyan:2018gqx}, which only evaluated correlated changes in the FSR and ISR PS scales.  
\item \textit{Top quark \pt:} Recent calculations suggest that the top quark \pt spectrum is strongly affected by NNLO effects~\cite{Czakon:2015owf,Czakon:2017wor,Catani:2019hip}. The \pt of the top quark in simulation is varied to match the distribution measured by CMS~\cite{Khachatryan:2016mnb, Sirunyan:2017mzl}.
The default simulation is not corrected for this effect, but this variation is included via a nuisance parameter in the  \mtop measurement.
\item \textit{Underlying event:} Measurements of the underlying event have been used to tune \PYTHIA parameters describing nonperturbative QCD effects~\cite{Skands:2014pea, CMS:2019csb}.
  The parameters of the tune are varied within their uncertainties.
\item \textit{Early resonance decays:}
  Modeling of color reconnection introduces systematic uncertainties, which are estimated by comparing different CR models and settings.
  In the default sample, the top quark decay products are not included in the CR process.
  This setting is compared to the case of including the decay products by enabling early resonance decays in \PYTHIAeight.
\item \textit{CR modeling:}
  In addition to the default model used in \PYTHIAeight, two alternative CR models are used, namely
  a model with string formation beyond leading color (``QCD inspired'')~\cite{Christiansen:2015yqa}
  and a model allowing the gluons to be moved to another string (``gluon move'')~\cite{Argyropoulos:2014zoa}.
  Underlying event measurements are used to tune the parameters of all models~\cite{CMS:2022awf}.
  For each model, an individual nuisance parameter is introduced.
\end{itemize}
\par}

\section{Mass extraction method}
A maximum likelihood (ML) fit to the selected events is employed to measure \mtop. 
The evaluated likelihood ratio  $\lambda(\mtop, \vectheta, \vecbeta, \Omega | \text{data})$ depends not only on \mtop, but also on three sets of nuisance parameters.
The nuisance parameters,  \vectheta, incorporate the uncertainty in systematic effects, while the statistical nuisance parameters, \vecbeta and $\Omega$, incorporate the statistical uncertainties in the simulation.
The  parameters \vecbeta account for the limited sample size of the default simulation and the parameters $\Omega$ account for the limited size in the simulated samples for variations of \mtop and of the uncertainty sources.
All nuisance parameters are normalized such that a value of 0 represents the absence of the systematic effect and the values $\pm$1 correspond to a variation of the systematic effect by one standard deviation up or down. 
 The \textsc{RooFit}~\cite{Verkerke:2003ir} package is used to define and evaluate all the functions. The minimum of the negative log-likelihood  $-2 \ln \lambda(\mtop, \vectheta, \vecbeta, \Omega| \text{data})$ is found with the \textsc{Minuit2} package~\cite{James:1975dr}.

The data are characterized by up to four observables per event, as mentioned in Section~\ref{sec:obs}.
The events are split into the electron+jets and the muon+jets channels.
The input to the ML fit is a set of one-dimensional histograms of the observables, $x_i$, in the two \Pgof categories. 
For each histogram, a suitable probability density function $P(x_i |\mtop, \vectheta, \vecbeta, \Omega)$ is derived from simulation.

The probability density function for the \mtfit histograms is approximated by the sum of a Voigt profile  (the convolution of a Cauchy--Lorentz distribution and a Gaussian distribution) for the correctly reconstructed \ttbar candidates and Chebyshev polynomials for the remaining events. 
For all other observables, a binned probability density function  is used that returns the relative fraction of events per histogram bin.
Here,  eight bins are used for each observable and the width of the bins is chosen so that each bin has a similar number of selected events for the default simulation ($\mtgen = 172.5\GeV$). 
For the following, we denote the parameters of the probability density functions as $\vecalpha$.
All the functions $P_i(x_i | \vecalpha)$ are normalized to the number of events in the histogram for the observable $x_i$, so only shape information is used in the ML fit. 
Hence, the parameters $\vecalpha$ are correlated even for the  binned probability density function.
The dependence of these parameters on \mtop and \vectheta is assumed to be linear.   
The full expression is for a component $\alpha_k$ of \vecalpha
\begin{linenomath}
 \begin{equation*}
  \ifthenelse{\boolean{cms@external}}
{
\begin{aligned}
\alpha_k =\ & C_k \Bigl(1 + d_k \   \Bigl[ \alpha_k^0  + \sigma^\alpha_k \ \beta_k +  s_k^0 \  (\mtop -172.5\GeV)  \\
 &  + \sigma_k^0 \Omega_k^0 \Bigr]\Bigr)  \   \prod_l \Bigl(1+ d_k \  \Bigl[s_k^l \  \theta_l + \sigma_k^l \ \Omega_k^l  \Bigr] \Bigr),
  \end{aligned}
 }
 {
\alpha_k =  C_k \left(1 + d_k \   \left[ \alpha_k^0  + \sigma^\alpha_k \ \beta_k +  s_k^0 \  (\mtop -172.5\GeV) + \sigma_k^0 \ \Omega_k^0\right]\right)
 \   \prod_l \left(1+ d_k \  \left[s_k^l \  \theta_l + \sigma_k^l \ \Omega_k^l  \right] \right),
 }
 \end{equation*}
\end{linenomath}
with  $l$ indicating the nuisance parameter.
For the nuisance parameters corresponding to the FSR PS scale variations, the linear term,  $s_k^l \  \theta_l$,  is replaced with a second-order polynomial.
With these expressions for the parameters $\alpha_k$, the probability density function, $P_i(x_i | \vecalpha)$,  for an observable $x_i$ becomes the function $P_i(x_i | \mtop, \vectheta,  \vecbeta, \Omega)$ mentioned above.

The model parameter $\alpha_k^0$ is determined by a fit to the default simulation, while the linear dependencies of $ \alpha_k$ on  \mtop or a component $ \theta_l$ of \vectheta are expressed with the model parameters $s_k^0$ and $s_k^l$, respectively. 
The parameter $s_k^0$  is determined  from a simultaneous fit to simulated samples, where \mtgen is varied by $\pm$3\GeV from the default value. 
Along the same lines, the parameters  $s_k^l$  are obtained from fits to the simulation of the systematic effect corresponding to the nuisance parameter $\theta_l$.
The values of $C_k$ and $d_k$ are chosen ad hoc so that the results of the fits of $\alpha_k^0$, $s_k^0$, and the $s_k^l$ are all of the same order of magnitude and with a similar statistical uncertainty. This improves the numerical stability of the final ML fit.

The limited size of the simulated samples for different \mtgen values gives rise to a calibration uncertainty in \mtop. 
Hence, additional  statistical nuisance parameters, $\beta_k$ and  $\Omega_k^0$, are introduced that account for the statistical uncertainty in the 
model parameters $\alpha_k^0$ and $s_k^0$, similar to the Barlow--Beeston approach~\cite{Barlow:1993dm, Conway:2011in}. 
They are scaled by  $\sigma^\alpha_k$ and  $\sigma_k^0$, which are the standard deviations of $\alpha_k^0$ and $s_k^0$ obtained from the fits to the default simulation or the \mtgen-varied samples, respectively.
Hence, a  statistical nuisance with value $\Omega_k^0=\pm1$ changes the corresponding $\alpha_k$, i.e., the shape or the bin content of an observable, by the statistical uncertainty in the \mtop dependence evaluated at a shift in \mtop of 1\GeV. 
Similarly, the parameters  $s_k^l$ contain random fluctuations if they are determined from simulated samples that are statistically independent to the default simulation and of limited size.
These fluctuations can lead to overconstraints on the corresponding nuisance parameter $\theta_l$ and, hence, an underestimation of the systematic uncertainty.
The nuisance parameters $\Omega_k^l$  are added to counter these effects and are scaled by parameters $\sigma_k^l$, which are the standard deviations of the $s_k^l$ parameters from the fits to the corresponding samples for the systematic effect.
As in the \mtop case, a value of $\Omega_k^l=\pm1$ changes the corresponding $\alpha_k$ by the statistical uncertainty in the  $\theta_l$ dependence evaluated at a shift in $\theta_l$ of 1.
Unlike the systematic nuisance parameters $\theta_l$, which affect all  $\alpha_k$ collectively,  for each $\alpha_k$ there are individual $\Omega_k^l$ parameters. While this drastically increases the number of fitted parameters in the ML fit to data, this also guarantees that the  $\Omega_k^l$ parameters are hardly constrained by the fit to data and the uncertainty in \mtop includes the statistical uncertainty in the impact of the systematic effect. 

{\tolerance=1500
For a single histogram in a set, the products of Poisson probabilities for the prediction $\mu_{i,j} = n_{\text{tot},i}P_i(x_{i,j}|\mtop, \vectheta, \vecbeta, \Omega)$ and for an alternative model with an adjustable parameter per bin $\hat \mu_{i,j}  = n_{i,j}$ are used to compute the likelihood ratio $\lambda_i$~\cite{PDG2020}, where  $x_i$ is the observable, $n_{i,j}$ is the content of bin $j$ with bin center $x_{i,j}$, and $n_{\text{tot},i}$ is  the total number of entries.
Then the combined likelihood ratio for a set with observables $\vec{x}$ is 
\begin{linenomath}
\begin{equation*}
\label{eq:likeli}
  \ifthenelse{\boolean{cms@external}}
{
\begin{aligned}
\lambda(\mtop, \vectheta, \vecbeta, \Omega | \text{data}) = & \left(\prod_i \lambda_i(\mtop, \vectheta, \vecbeta, \Omega | x_i)\right)\\
& \times \left(\prod_l P(\theta_l)\right) P(\vecbeta) P(\Omega)\text{, }
\end{aligned}
}
{
\lambda(\mtop, \vectheta, \vecbeta, \Omega | \text{data}) = \left(\prod_i \lambda_i(\mtop, \vectheta, \vecbeta, \Omega | x_i) \right) \left(\prod_l P(\theta_l)\right) P(\vecbeta) P(\Omega)\text{, }
}
\end{equation*}  
\end{linenomath}
where $P(\theta_l)$, $P(\vecbeta)$, and $P(\Omega)$ are the pre-fit probability density functions of the nuisance parameters $\theta_l$, $\vecbeta$, and $\Omega$.
 The product of the likelihood ratios can be used on the right-hand side of the equation, as all observables are independent in most phase space regions.
The probability density functions of the nuisance parameters related to the sources of systematic uncertainties, $P(\theta_l)$, are standard normal distributions. 
All model parameters $\alpha_k^0$, $s_k^0$, or $s_k^l$ that are related to the same observable and nuisance parameter are obtained together by one fit to the corresponding simulation samples.
To take the correlations between these parameters into account,  the  statistical  nuisance parameters  $\vecbeta$ or $\Omega$ that incorporate the statistical uncertainty in these parameters are constrained by centered multivariate normal distributions. 
The covariances of the distributions are set  to the correlation matrices obtained by the fits of the corresponding parameters.
The latter nuisance parameters and constraints are only included if the model parameters are determined from samples that are statistically independent of the default simulation, like, for example, for the alternative color reconnection models.
If the model parameters are determined from samples obtained from varying the default simulation with event-based weights or scaling or smearing of the jet energies, the corresponding $\Omega$ parameters are fixed to zero and the constraint is removed from $\lambda(\mtop, \vectheta, \vecbeta, \Omega | \text{data})$.
\par}

The mass of the top quark is determined  with the profile likelihood fit for different sets of data histograms. The sets and their labels  are listed in Table~\ref{tab:obs}.

\begin{figure}[ht]
\noindent \centering
\includegraphics[width=0.45\textwidth]{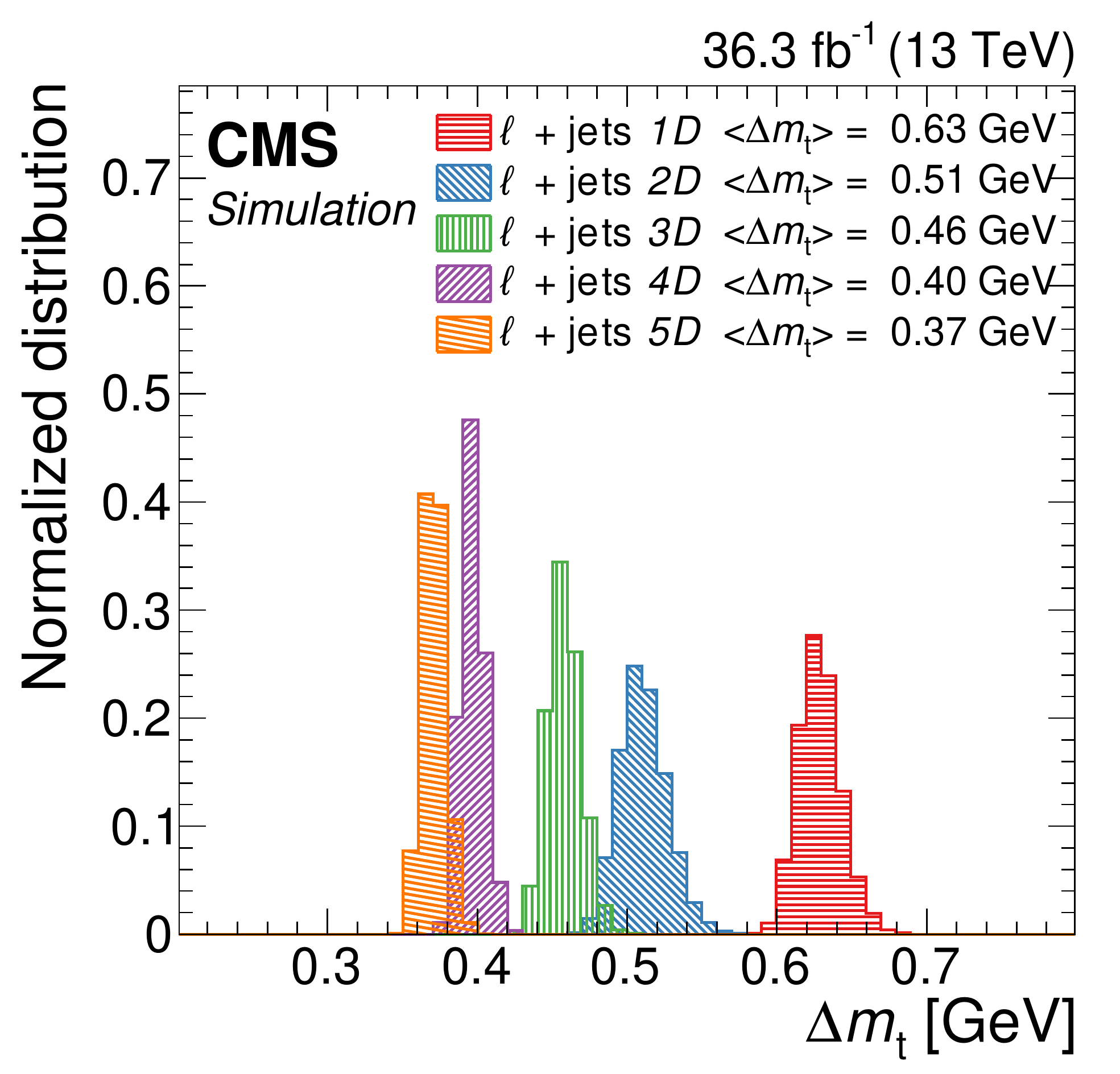}
\includegraphics[width=0.45\textwidth]{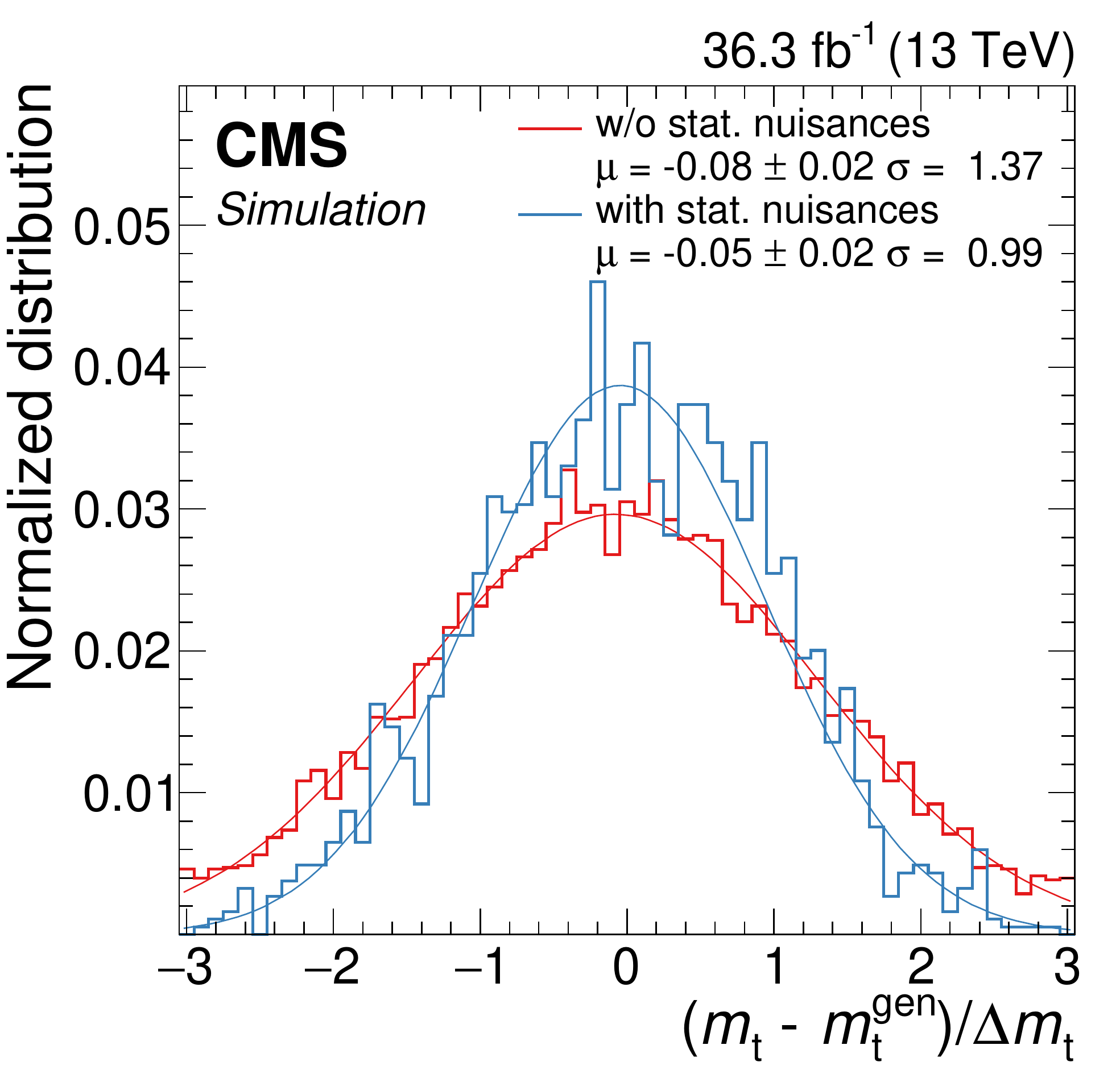}

\caption{\label{figpull} \label{fig:tots}
\CMSLeft:  Comparison of the expected total uncertainty in \mtop in the combined lepton+jets channel and for the different observable-category sets defined in Table~\ref{tab:obs}.
\CMSRight: The difference between the measured and generated \mtop values, divided by the uncertainty reported by the fit from pseudo-experiments without (red) or with (blue) the statistical nuisance parameters  \vecbeta and  $\Omega$  in the \textit{5D} ML fit.
Also included in the legend are the $\mu$ and $\sigma$ parameters of Gaussian functions (red and blue lines) fit to the histograms.
}
\end{figure}

The expected total uncertainty in \mtop is evaluated for each set defined in Table~\ref{tab:obs} with pseudo-experiments using the default simulation.
The results of the pseudo-experiments are shown in Fig.~\ref{fig:tots} (\cmsLeft). 
The improvements in the data reconstruction and calibration, event selection, simulation, and mass extraction method reduce the uncertainty in  the \textit{1D} measurement from 1.09 to 0.63\GeV, when compared to the previous measurement~\cite{Sirunyan:2018gqx}.
The uncertainty in the \textit{2D} measurement improves from 0.63 to 0.50\GeV.
The additional observables and the split into categories further reduce the expected uncertainty down to 0.37\GeV for the \textit{5D} set.

The statistical uncertainty is obtained from fits that only have \mtop as a free parameter.
From studies on simulation, it is expected to be 0.07, 0.06, and 0.04\GeV in the electron+jets, muon+jets, and the combined (lepton+jets) channels, respectively.

{\tolerance=800
The applied statistical model is verified with additional pseudo-experiments. 
Here, the data for one pseudo-experiment are generated using probability density functions $P(x_i |\mtop, \vectheta)$ that have the same functional form as the ones used in the ML fit, but their model parameters \vecalpha and all slopes, $s_k^l$ are determined on statistically fluctuated simulations.
For the generation of a pseudo-experiment, \mtop is chosen from a uniform distribution with a mean of 172.5\GeV and the same standard deviation as is assumed for the calibration uncertainty.
The values of the nuisance parameters \vectheta are drawn from standard normal distributions. 
The same ML fit that is applied to the collider data is then performed on the pseudo-data. 
The pseudo-experiments are generated for two cases, specifically, with and without  the  statistical nuisance parameters  \vecbeta and  $\Omega$  in the ML fit.
Figure~\ref{figpull} (\cmsRight) shows the distribution of the differences between the measured and generated \mtop values, divided by the uncertainty reported by the fit for both cases.
A nearly 40\% underestimation of the measurement uncertainty can be seen for the case without  the statistical nuisance parameters  \vecbeta and  $\Omega$, while consistency is observed for the method that is employed on data.
\par}

In addition, single-parameter fits were performed on pseudo-data sampled from simulation to verify that the mass extraction method is unbiased and reports the correct uncertainty. 
These tests were done for fits of \mtop with samples corresponding to mass values of 169.5, 172.5, and 175.5\GeV, as well as on the simulation of different systematic effects for the fits of the corresponding nuisance parameter.

\section{Results}
\label{sec:results}

\begin{figure}[ht]
\noindent \centering
\includegraphics[width=0.45\textwidth]{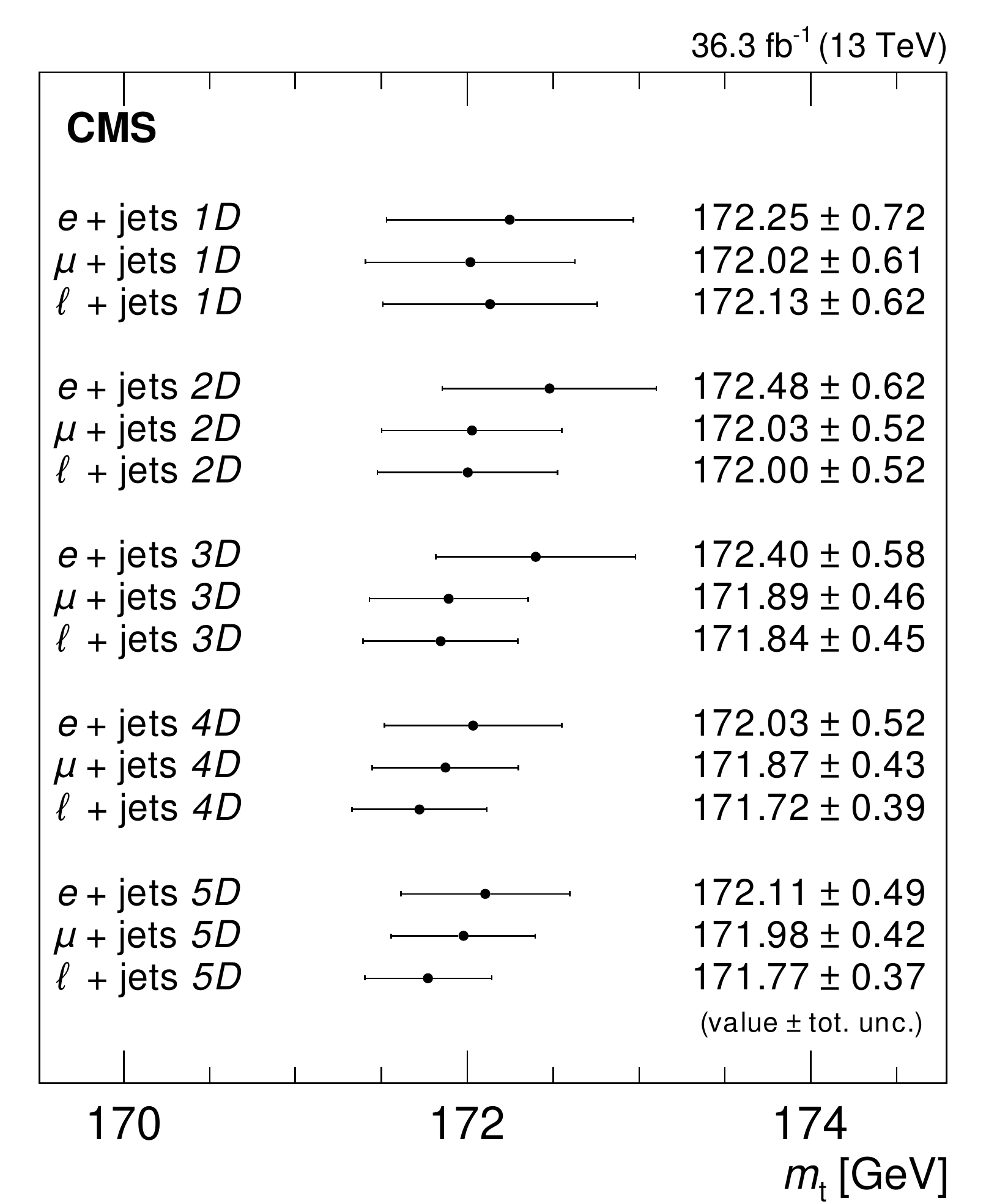}

\caption{\label{fig:results}
Measurement of \mtop in the three different channels for the different sets of observables and categories as defined in Table~\ref{tab:obs}. 
}
\end{figure}

{\tolerance=800
The results of the  profile likelihood fits to data are shown in Fig.~\ref{fig:results} for the electron+jets, muon+jets, and lepton+jets channels and for the different sets of observables and categories, as defined in Table~\ref{tab:obs}.
The observables \mWreco, $\mlb/\mtfit$, and \Rb provide constraints on the modeling of the \ttbar decays in addition to the observables \mtfit and $\mlb|_{\Pgof<0.2}$, which are highly sensitive to \mtop.
With the profile likelihood method, these constraints not only reduce the uncertainty  in \mtop, but  also change the measured \mtop value, as they effectively alter the parameters of the reference \ttbar simulation. 
When additional observables are included, the measurement in the lepton+jets channel yields a smaller mass value than the separate channels because of  the correlations between the channels.
\par}

\begin{figure*}[ht]
  \noindent
  \centering
  \includegraphics[width=0.42\textwidth]{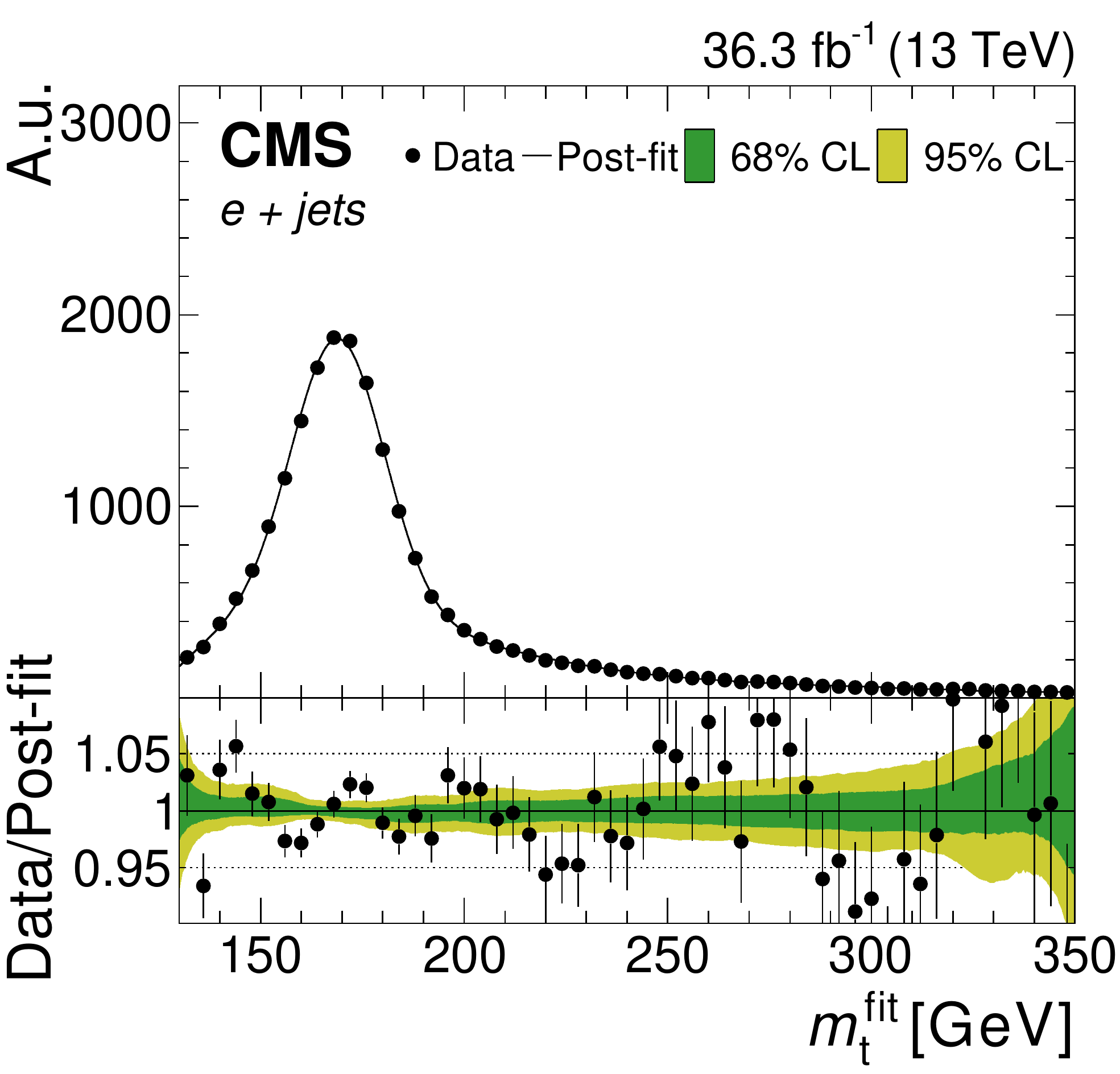}
  \includegraphics[width=0.42\textwidth]{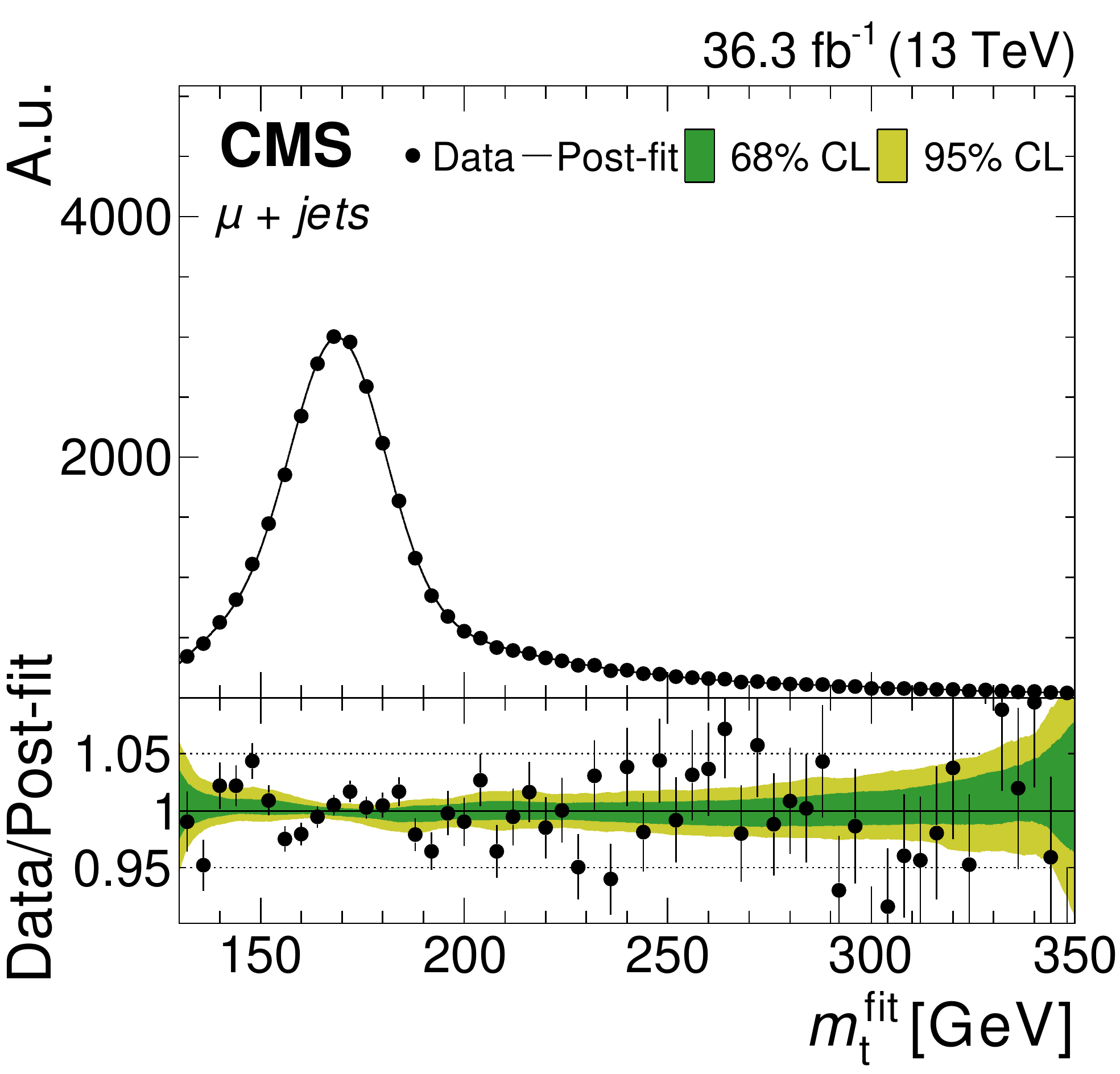}\\
  \includegraphics[width=0.42\textwidth]{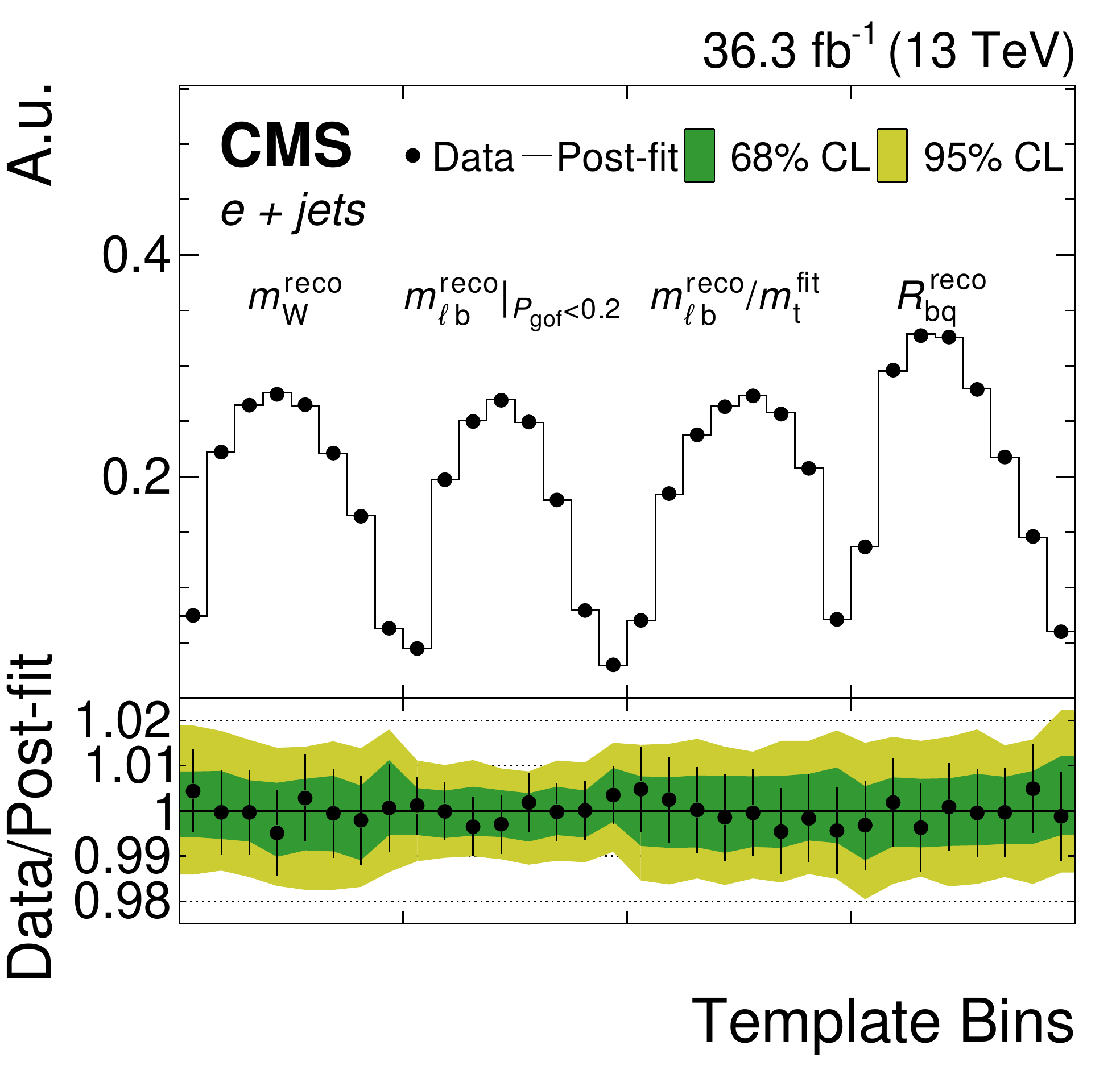}
  \includegraphics[width=0.42\textwidth]{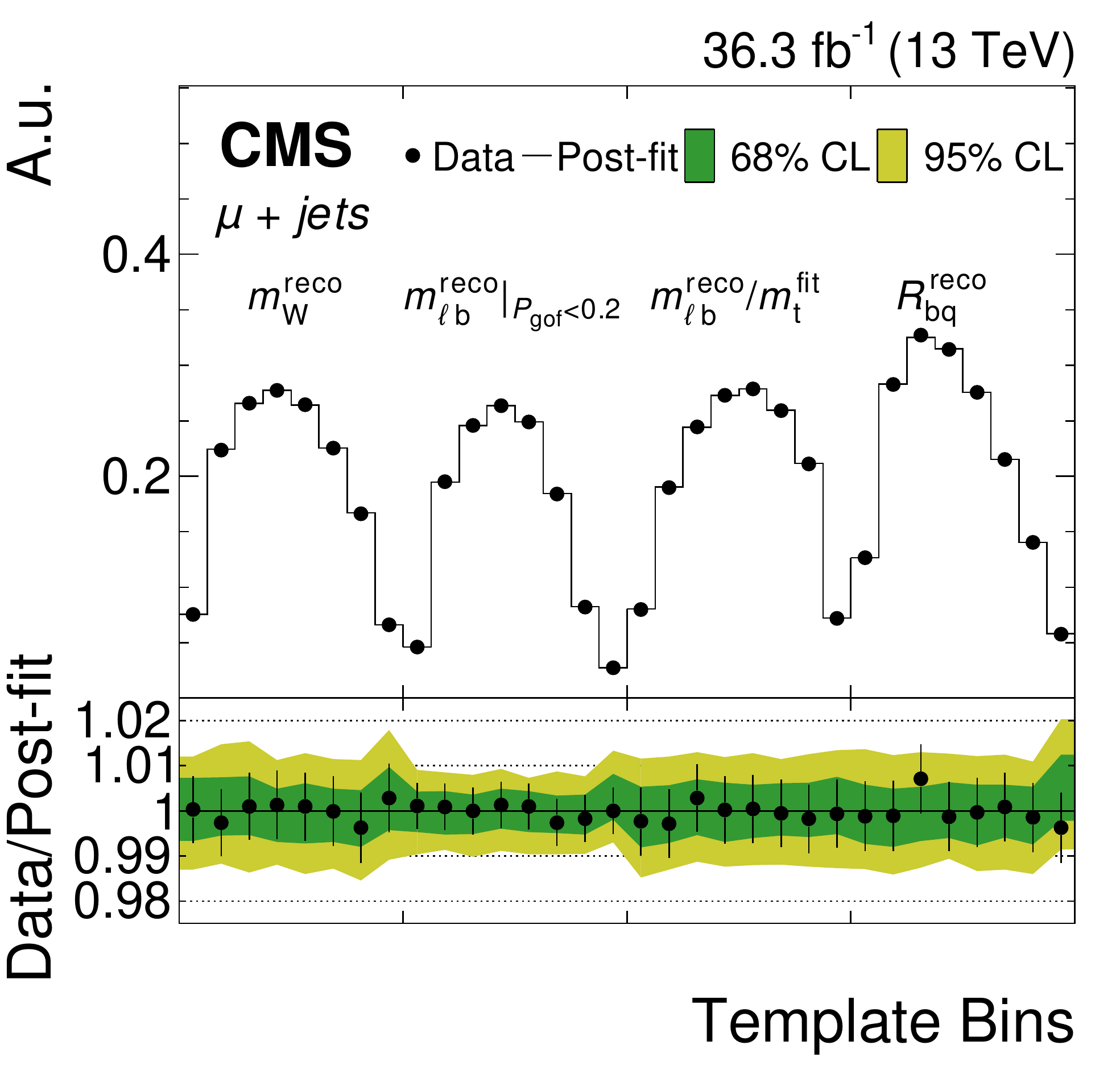}
\caption{\label{fig:post-fit}
Distribution of \mtfit (upper) and the additional observables (lower) that are the input to the \textit{5D} ML fit and their post-fit probability density functions for the combined fit to the electron+jets (left) and muon+jets (right) channels.  For the additional observables results,  the events in each bin are divided by the bin width. The lower panels show the ratio of data and post-fit template values. The green and yellow bands represent the 68 and 95\% confidence levels in the fit uncertainty.}
\end{figure*}

The \textit{5D} fit to the selected events results in the best precision and yields in the respective channels:
\begin{eqnarray*}
\text{Electron+jets:} & \mtop^{\textit{5D}} & =  172.11\pm0.49 \GeV,\\
\text{Muon+jets:} & \mtop^{\textit{5D}} & =  171.98\pm0.42 \GeV,\\
\text{Lepton+jets:} & \mtop^{\textit{5D}} & =  171.77\pm0.37 \GeV.
\end{eqnarray*}
The comparisons of the data distributions and the post-fit \textit{5D} model are shown in Fig.~\ref{fig:post-fit}.
While the binned probability density functions of the model describe the corresponding observables well,
significant deviations between the data and the model can be observed in the peak region of the \mtfit observable.
These deviations are also observed in simulation and stem from the fact that effectively only two parameters, the peak position and its average width, are used in the model to describe the peak. 
Tests with simulation show no bias on the extracted \mtop value from these deviations.
In fact, the small number of parameters should increase the robustness of the measurement as the model is not sensitive to finer details of the peak shape that might be difficult to simulate correctly.

\begin{figure*}[ht]
\noindent \centering
\includegraphics[width=\textwidth]{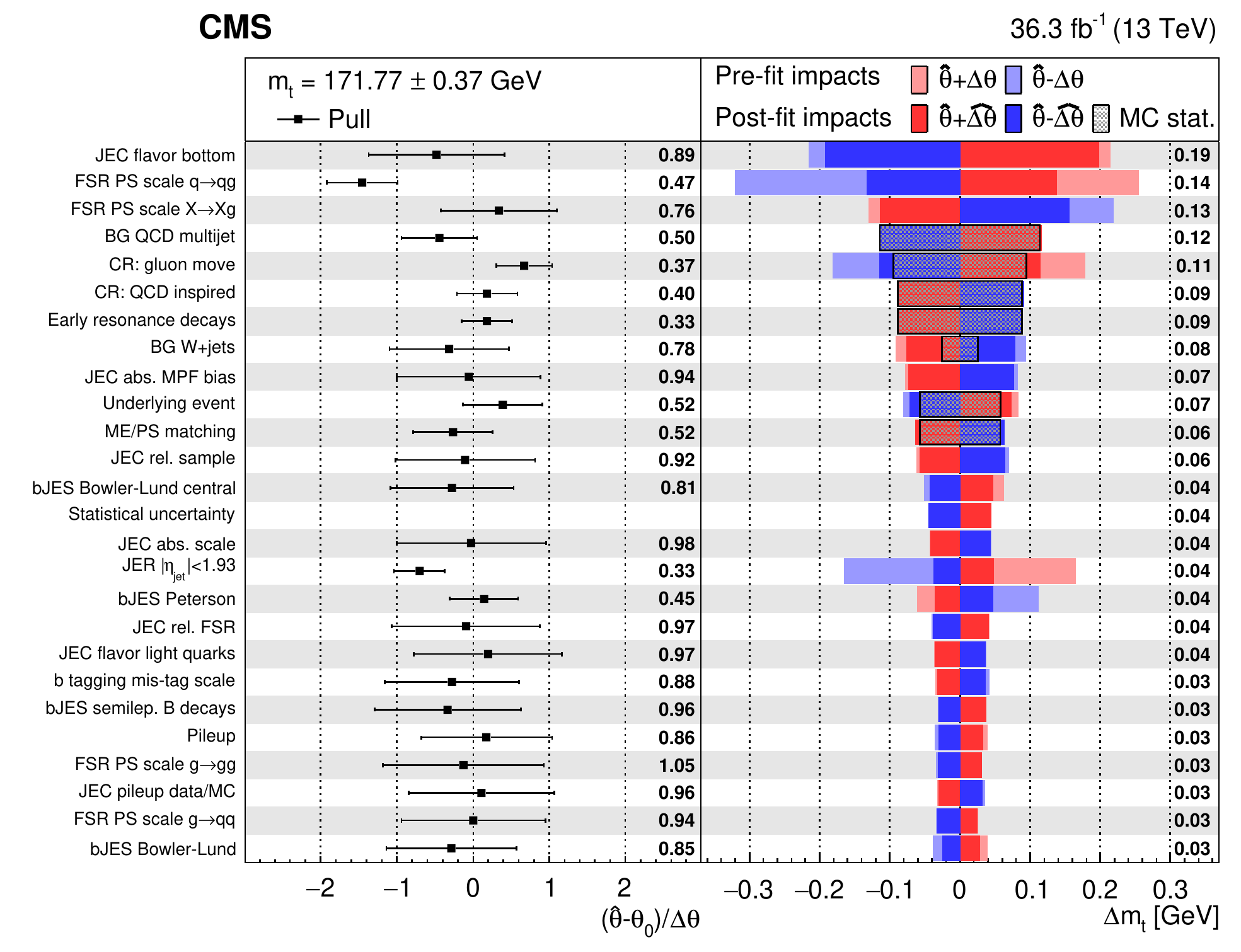}
\caption{\label{fig:res5D}
Measurement of \mtop in the combined lepton+jets channel using the \textit{5D}  set of observables and categories.
The left plot shows the post-fit pulls on the most important nuisance parameters and the numbers quote the post-fit uncertainty in the nuisance parameter. 
The right plot shows their pre-fit (lighter colored bars) and post-fit impacts (darker colored bars) on \mtop for up (red) and down (blue) variations. 
The post-fit impacts of systematic effects that are affected by the limited size of simulation samples include the contribution from the additional statistical nuisance parameters accounting for the effect. 
The size of the additional contribution from the statistical nuisance parameters is called  MC stat. and shown as gray-dotted areas. 
The average of the post-fit impacts in \GeV for up and down variations is printed on the right. 
The rows are sorted by the size of the averaged post-fit impact. 
The statistical uncertainty in \mtop is depicted in the corresponding row.
}
\end{figure*}
Figure~\ref{fig:res5D} shows the pulls on the most important systematic nuisance parameters $\theta$ and their impacts on \mtop, $\Delta\mtop$, after the fit with the \textit{5D} model.
The pulls are defined as $(\hat \theta- \theta_0)/\Delta\theta$, where $\hat \theta$ is the measured nuisance parameter value and $\theta_0$ and $\Delta\theta$ are the mean and standard deviation of the nuisance parameter before the fit.
The pre-fit impacts are evaluated by repeating the ML fit with the studied nuisance parameter $\theta$ fixed to $\hat \theta \pm \Delta \theta$ and taking the difference in \mtop between the result of these fits and the measured \mtop.
In most cases, the post-fit impacts are evaluated respectively with  $\hat \theta$ and $ \widehat{\Delta \theta}$, where $\widehat{\Delta \theta}$ is the uncertainty in the nuisance parameter after the fit.
However, if the studied systematic nuisance parameter $\theta$ has statistical nuisance parameters $\Omega$ that account for the statistical uncertainty in the $\theta$-dependence of the model, 
 the combined impact of the systematic and statistical nuisance parameters is plotted in Fig.~\ref{fig:res5D} as post-fit impact.
To estimate this combined impact, the likelihood fit is repeated with the corresponding systematic and statistical nuisance parameters fixed to their post-fit values and the quadratic difference of the overall \mtop uncertainty compared to the default fit is taken.  
The quadratic difference between the combined impact and the post-fit impact of only the systematic nuisance parameter is interpreted as the effect of the limited size of the systematic simulation samples.

Most nuisance parameters are consistent with their pre-fit values. 
The largest effect on the measured mass value corresponds to the FSR scale of the  $\PQq\to \PQq\Pg$ branching type. 
The effect is caused by the difference in the peak position of \mWreco seen in Fig.~\ref{fig3} (\cmsLeft). 
The previous measurements in this channel by the CMS Collaboration assumed correlated FSR PS scales with the same scale choice for jets induced by light quarks and  \PQb quarks~\cite{Khachatryan:2015hba,Sirunyan:2018gqx}.
In that case, a lower peak position  in the \mWreco distribution would also cause the \mtfit peak position to be lower than expected from simulation for a given \mtop value, resulting in a higher top quark mass value to be measured.
In fact, a \textit{5D} fit to data assuming fully correlated FSR PS scale choices yields $\mtop =  172.20 \pm 0.31\GeV$.
This value is very close to the previous measurement on the same data of $\mtop = 172.25 \pm 0.63\GeV$~\cite{Sirunyan:2018gqx}.

\begin{figure}[ht]
\noindent \centering
\includegraphics[width=0.45\textwidth]{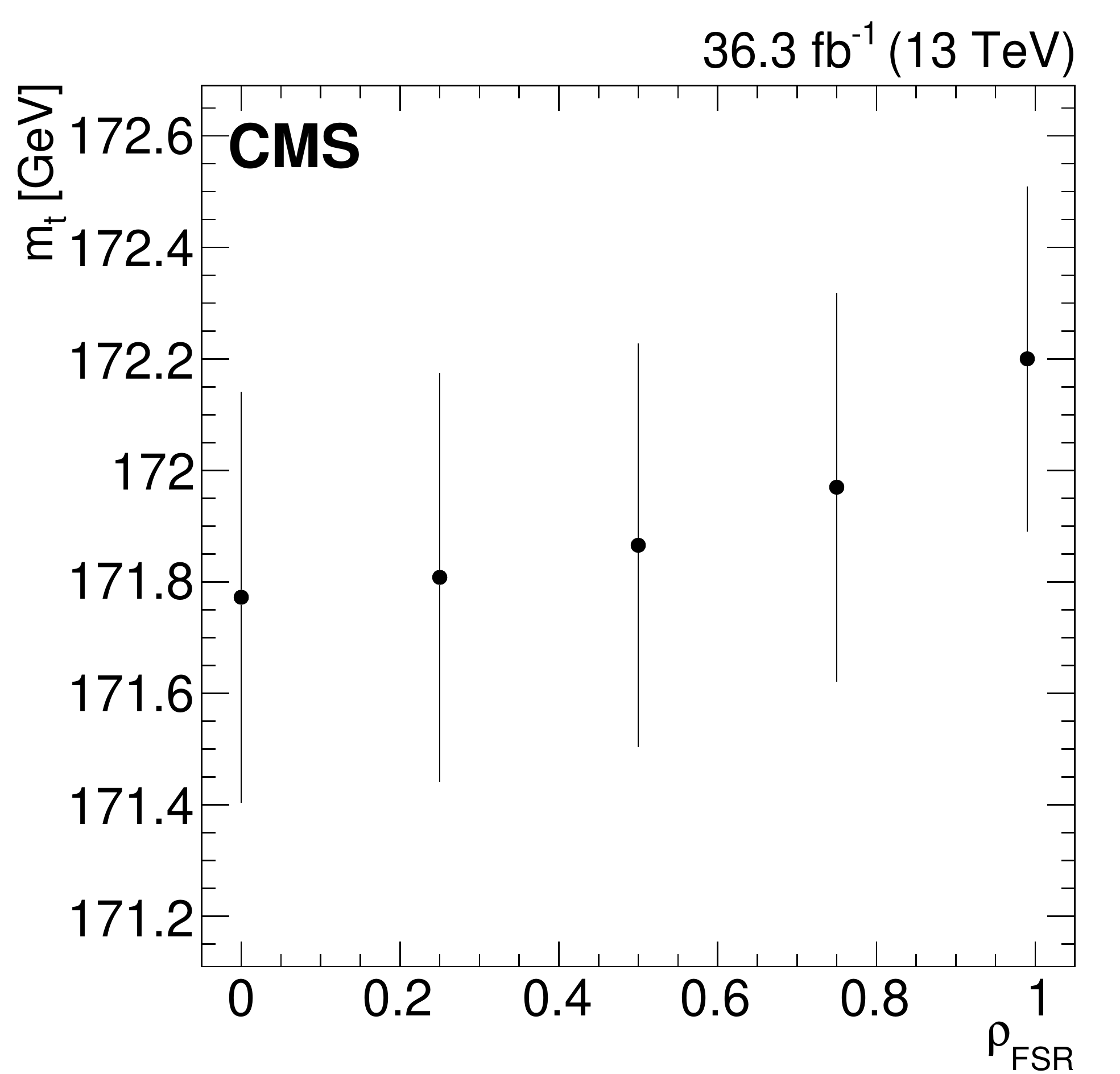}
\caption{\label{fig:resultsfsr}
Dependence of the  \textit{5D} result on the assumed correlation $\rho_\text{{FSR}}$ between the  FSR PS scales in the lepton+jets channel.}
\end{figure}

The measurement is repeated for different correlation coefficients ($\rho_\text{{FSR}}$) in the pre-fit covariance matrix between the FSR PS scales for the different branching types.
The result of this study is shown in Fig.~\ref{fig:resultsfsr}. 
The final result strongly depends on the choice of the correlation coefficient between the FSR PS scales because of the significant deviation for the FSR PS scale of the  $\PQq\to \PQq\Pg$ branching from the default simulation. 
However, the assumption of  strongly correlated  FSR PS scale choices would also significantly reduce the overall uncertainty, as the impacts from the scale choice for gluon radiation from \PQb quarks ($\PX\to \PX\Pg$) and light quarks ($\PQq\to \PQq\Pg$) partially cancel.
In addition, there is a tension between the measured nuisance parameter values for the different FSR PS scales, which disfavors a strong correlation. 
As there is only a small dependence on  FSR PS scale correlations at low correlation coefficients ($\rho_\text{{FSR}} < 0.5$),
and uncorrelated  nuisance parameters for the FSR PS scales  receive  the least constraint from the fit to data, 
we assume  uncorrelated FSR PS scales for this measurement.

Most of the other nuisance parameters that show a strong post-fit constraint correspond to systematic uncertainties that are evaluated on independent samples of limited size.
The small sample sizes are expected to bias these nuisances parameter and lead to too small uncertainties.
Hence, the nuisance paramerts are accompanied by additional statistical nuisance parameters.
A comparison of the pre-fit and post-fit impacts where the post-fit impacts include the impact of these statistical nuisance parameters shows that there is an only minimal constraint by the fit on the corresponding systematic uncertainties. 

{\tolerance=800
The largest constraint of a nuisance parameter without additional statistical nuisance parameters corresponds to the JER uncertainty.
This is expected, as the energy resolution of jets from \ttbar decays can be measured much better from the width of the \mWreco distribution than by the extrapolation of the resolution measurement with dijet topologies at much higher transverse momenta~\cite{Khachatryan:2016kdb}. 
\par}

\begin{table}[!htb]
  \topcaption{Comparison of the uncertainty in the top quark mass in the previous measurement~\cite{Sirunyan:2018gqx, Sirunyan:2018mlv} and the new \textit{2D} and \textit{5D} results  in the lepton+jets channel.    
  \label{tab:syst_comp}}
  \centering
  \begin{tabular}{ld{1.2}d{1.2}d{1.2}}
      & \multicolumn{3}{C}{\delta \mtop~\text{[\GeVns]}}\\
      &\multicolumn{1}{c}{previous \textit{2D}}&\multicolumn{1}{c}{\textit{2D}}&\multicolumn{1}{c}{\textit{5D}}\\\hline
    \multicolumn{4}{l}{\textsl{Experimental uncertainties}}\\
    Method calibration                  &  0.05 & 0.02&  0.02 \\
    JEC 			                 &  0.18 & 0.32&  0.16 \\
    -- Intercalibration                 & 0.04 & 0.10 &  0.04 \\
    -- MPFInSitu                        & 0.07 &  0.15 & 0.07 \\
    -- Uncorrelated                     & 0.16 & 0.21 & 0.10 \\
    Jet energy resolution               & 0.12& 0.12& 0.05\\
    \PQb~tagging                       &  0.03  & 0.01&  0.03 \\
    Lepton SFs and mom. scale   &  & 0.00 & 0.03\\
    Pileup                              & 0.05 &  0.00 & 0.03 \\
    Background                  & 0.02 & 0.12& 0.15\\
      [\cmsTabSkip]
    \multicolumn{4}{l}{\textsl{Modeling uncertainties}}\\
    JEC flavor             & 0.39 & 0.30& 0.20 \\
    \PQb-jet modeling       &  0.12 & 0.15&  0.11 \\
     PDF                                 &  0.02 & 0.00 &  0.01 \\
    Renorm. and fact. scales               &  0.01 & 0.03 &  0.02 \\
    ME/PS matching                      & 0.07 & 0.06 &  0.07\\
    ME generator                        & 0.20 & \NA& \NA \\
    ISR PS scale                        & 0.07 & 0.01 & 0.01 \\
    FSR PS scale                        & 0.13 & 0.37 & 0.21 \\
    Top quark \pt                        & 0.01 & 0.06 & 0.00 \\
    Underlying event                    & 0.07 &0.09& 0.04 \\
    Early resonance decays              & 0.07 &0.13& 0.09 \\
    CR modeling          & 0.31 & 0.15& 0.15 \\
    [\cmsTabSkip]
    Statistical             &  0.08 & 0.05&  0.04 \\
     [\cmsTabSkip]
    Total                   &  0.63 & 0.52&  0.37 \\
  \end{tabular}
\end{table}
{\tolerance=800
Table~\ref{tab:syst_comp} compares the measurements by the \textit{2D} and \textit{5D} methods with the previous result~\cite{Sirunyan:2018gqx, Sirunyan:2018mlv} for the same data-taking period. 
The JEC uncertainties are grouped following the recommendations documented in Ref.~\cite{CMS-PAS-JME-15-001}.
The uncertainty in \mtop for one source (row) in this table is evaluated from the covariance matrix of the ML fit by taking the square root of  $\text{cov}(\mtop,X) \text{cov}(X,X)^{-1} \text{cov}(X,\mtop)$, where $\text{cov}(\mtop,X)$, $\text{cov}(X,X)$, $\text{cov}(X,\mtop)$ are the parts of the covariance matrix related to \mtop or the set of nuisance parameters $X$ contributing to the source, respectively. 
The statistical and calibration uncertainties are obtained  differently by computing the partial covariance matrix on \mtop where all other nuisance parameters are removed.
The quadratic sum of all computed systematic uncertainties is larger than the uncertainty in \mtop from the ML fit, as the sum ignores the post-fit correlations between the systematic uncertainty sources. 
\par}

The  \textit{5D} method is the only method that surpasses the strong reduction in the uncertainty in the JEC achieved by the previous analysis that determined \mtop and in situ an overall jet energy scale factor (JSF). 
However, the measurement presented here also constrains the jet energy resolution uncertainty that was unaffected by the JSF.
The new observables and additional events with a low \Pgof reduce most modeling uncertainties, but lead to a slight increase in some experimental uncertainties.
While the usage of weights for the PS variations removes the previously significant statistical component in the PS uncertainties, the introduction of separate PS scales leads to a large increase in the uncertainty in the FSR PS scale, despite the tight constraint on the corresponding nuisance parameters shown in Fig.~\ref{fig:res5D}.
  
The result presented here achieves a considerable improvement compared to all previously published top quark mass measurements.
Hence, it supersedes the previously published measurement in this channel on the same data set~\cite{Sirunyan:2018gqx}.
The analysis shows the precision that is achievable from direct measurements of the top quark mass. 
As the uncertainty in the relationship of the direct measurement from simulation templates to a theoretically well-defined top quark mass is currently of similar size, the measurement should fuel further theoretical studies on the topic.

\section{Summary}
The mass of the top quark is measured using LHC proton-proton collision data collected in 2016 with the CMS detector at $\sqrt{s}=13\TeV$, corresponding to an integrated luminosity of  36.3\fbinv.
The measurement uses a sample of \ttbar events containing one isolated electron or muon and at least four jets in the final state.
For each event, the mass is reconstructed from a kinematic fit of the decay products to a \ttbar hypothesis.
A  likelihood method  is applied using up to four observables per event to extract the top quark mass and constrain the influences of systematic effects, which  are included as nuisance parameters in the likelihood.
The top quark mass is measured to be $171.77\pm 0.37\GeV$. This result achieves a considerable improvement compared to all previously published top quark mass measurements and supersedes the previously published measurement in this channel on the same data set.

\begin{acknowledgments}
We congratulate our colleagues in the CERN accelerator departments for the excellent performance of the LHC and thank the technical and administrative staffs at CERN and at other CMS institutes for their contributions to the success of the CMS effort. In addition, we gratefully acknowledge the computing centers and personnel of the Worldwide LHC Computing Grid and other centers for delivering so effectively the computing infrastructure essential to our analyses. Finally, we acknowledge the enduring support for the construction and operation of the LHC, the CMS detector, and the supporting computing infrastructure provided by the following funding agencies: BMBWF and FWF (Austria); FNRS and FWO (Belgium); CNPq, CAPES, FAPERJ, FAPERGS, and FAPESP (Brazil); MES and BNSF (Bulgaria); CERN; CAS, MoST, and NSFC (China); MINCIENCIAS (Colombia); MSES and CSF (Croatia); RIF (Cyprus); SENESCYT (Ecuador); MoER, ERC PUT and ERDF (Estonia); Academy of Finland, MEC, and HIP (Finland); CEA and CNRS/IN2P3 (France); BMBF, DFG, and HGF (Germany); GSRI (Greece); NKFIH (Hungary); DAE and DST (India); IPM (Iran); SFI (Ireland); INFN (Italy); MSIP and NRF (Republic of Korea); MES (Latvia); LAS (Lithuania); MOE and UM (Malaysia); BUAP, CINVESTAV, CONACYT, LNS, SEP, and UASLP-FAI (Mexico); MOS (Montenegro); MBIE (New Zealand); PAEC (Pakistan); MES and NSC (Poland); FCT (Portugal); JINR (Dubna); MON, RosAtom, RAS, RFBR, and NRC KI (Russia); MESTD (Serbia); MCIN/AEI and PCTI (Spain); MOSTR (Sri Lanka); Swiss Funding Agencies (Switzerland); MST (Taipei); MHESI and NSTDA (Thailand); TUBITAK and TENMAK (Turkey); NASU (Ukraine); STFC (United Kingdom); DOE and NSF (USA).
  
\hyphenation{Rachada-pisek} Individuals have received support from the Marie-Curie program and the European Research Council and Horizon 2020 Grant, contract Nos.\ 675440, 724704, 752730, 758316, 765710, 824093, 884104, and COST Action CA16108 (European Union); the Leventis Foundation; the Alfred P.\ Sloan Foundation; the Alexander von Humboldt Foundation; the Belgian Federal Science Policy Office; the Fonds pour la Formation \`a la Recherche dans l'Industrie et dans l'Agriculture (FRIA-Belgium); the Agentschap voor Innovatie door Wetenschap en Technologie (IWT-Belgium); the F.R.S.-FNRS and FWO (Belgium) under the ``Excellence of Science -- EOS" -- be.h project n.\ 30820817; the Beijing Municipal Science \& Technology Commission, No. Z191100007219010; the Ministry of Education, Youth and Sports (MEYS) of the Czech Republic; the Hellenic Foundation for Research and Innovation (HFRI), Project Number 2288 (Greece); the Deutsche Forschungsgemeinschaft (DFG), under Germany's Excellence Strategy -- EXC 2121 ``Quantum Universe" -- 390833306, and under project number 400140256 - GRK2497; the Hungarian Academy of Sciences, the New National Excellence Program - \'UNKP, the NKFIH research grants K 124845, K 124850, K 128713, K 128786, K 129058, K 131991, K 133046, K 138136, K 143460, K 143477, 2020-2.2.1-ED-2021-00181, and TKP2021-NKTA-64 (Hungary); the Council of Science and Industrial Research, India; the Latvian Council of Science; the Ministry of Education and Science, project no. 2022/WK/14, and the National Science Center, contracts Opus 2021/41/B/ST2/01369 and 2021/43/B/ST2/01552 (Poland); the Funda\c{c}\~ao para a Ci\^encia e a Tecnologia, grant CEECIND/01334/2018 (Portugal); the National Priorities Research Program by Qatar National Research Fund; the Ministry of Science and Higher Education, projects no. 0723-2020-0041 and no. FSWW-2020-0008 (Russia); MCIN/AEI/10.13039/501100011033, ERDF ``a way of making Europe", and the Programa Estatal de Fomento de la Investigaci{\'o}n Cient{\'i}fica y T{\'e}cnica de Excelencia Mar\'{\i}a de Maeztu, grant MDM-2017-0765 and Programa Severo Ochoa del Principado de Asturias (Spain); the Chulalongkorn Academic into Its 2nd Century Project Advancement Project, and the National Science, Research and Innovation Fund via the Program Management Unit for Human Resources \& Institutional Development, Research and Innovation, grant B05F650021 (Thailand); the Kavli Foundation; the Nvidia Corporation; the SuperMicro Corporation; the Welch Foundation, contract C-1845; and the Weston Havens Foundation (USA).
\end{acknowledgments}

\bibliography{auto_generated} \cleardoublepage \appendix\section{The CMS Collaboration \label{app:collab}}\begin{sloppypar}\hyphenpenalty=5000\widowpenalty=500\clubpenalty=5000
\cmsinstitute{Yerevan Physics Institute, Yerevan, Armenia}
{\tolerance=6000
A.~Tumasyan\cmsAuthorMark{1}\cmsorcid{0009-0000-0684-6742}
\par}
\cmsinstitute{Institut f\"{u}r Hochenergiephysik, Vienna, Austria}
{\tolerance=6000
W.~Adam\cmsorcid{0000-0001-9099-4341}, J.W.~Andrejkovic, T.~Bergauer\cmsorcid{0000-0002-5786-0293}, S.~Chatterjee\cmsorcid{0000-0003-2660-0349}, K.~Damanakis\cmsorcid{0000-0001-5389-2872}, M.~Dragicevic\cmsorcid{0000-0003-1967-6783}, A.~Escalante~Del~Valle\cmsorcid{0000-0002-9702-6359}, P.S.~Hussain\cmsorcid{0000-0002-4825-5278}, M.~Jeitler\cmsAuthorMark{2}\cmsorcid{0000-0002-5141-9560}, N.~Krammer\cmsorcid{0000-0002-0548-0985}, L.~Lechner\cmsorcid{0000-0002-3065-1141}, D.~Liko\cmsorcid{0000-0002-3380-473X}, I.~Mikulec\cmsorcid{0000-0003-0385-2746}, P.~Paulitsch, F.M.~Pitters, J.~Schieck\cmsAuthorMark{2}\cmsorcid{0000-0002-1058-8093}, R.~Sch\"{o}fbeck\cmsorcid{0000-0002-2332-8784}, D.~Schwarz\cmsorcid{0000-0002-3821-7331}, M.~Sonawane\cmsorcid{0000-0003-0510-7010}, S.~Templ\cmsorcid{0000-0003-3137-5692}, W.~Waltenberger\cmsorcid{0000-0002-6215-7228}, C.-E.~Wulz\cmsAuthorMark{2}\cmsorcid{0000-0001-9226-5812}
% ...
\par}
\vskip\cmsinstskip
\dag:~Deceased\\
$^{1}$Also at Yerevan State University, Yerevan, Armenia\\
$^{2}$Also at TU Wien, Vienna, Austria\\
% ...
\end{sloppypar}
\end{document}